\documentclass[draft]{agujournal2019}
\usepackage{url} %this package should fix any errors with URLs in refs.
\usepackage{lineno}
\usepackage[inline]{trackchanges} %for better track changes. finalnew option will compile document with changes incorporated.
\usepackage{soul}
\usepackage{amsmath}
\usepackage{longtable}
\usepackage{afterpage}
\draftfalse

\journalname{JGR: Planets}

\begin{document}

\title{What’s Inside Matters: The Effect of Oxygen Fugacity and Initial Volatile Abundance on the Atmospheres of the TRAPPIST-1 Planets}

\authors{Junellie Perez\affil{1,2}, Laura K. Schaefer\affil{2,3}, Edward Schwieterman\affil{2,4}, Kevin B. Stevenson\affil{2,5}, Howard Chen\affil{2,6}, Jacob Lustig-Yaeger\affil{2,5}}

\affiliation{1}{Johns Hopkins University, Baltimore, MD, USA}
\affiliation{2}{Consortium on Habitability and Atmospheres of M-dwarf Planets (CHAMPs), Laurel, MD, USA}
\affiliation{3}{Stanford University, Stanford, CA, USA}
\affiliation{4}{University of California, Riverside, Riverside, CA, USA}
\affiliation{5}{JHU Applied Physics Laboratory, Laurel, MD, USA}
\affiliation{6}{Florida Institute of Technology, Melbourne, FL, USA}

\correspondingauthor{Junellie Perez}{jgonza70@jhu.edu}

\begin{keypoints}

\item We tested the effect of interior parameters through the use of an atmosphere-interior exchange model on TRAPPIST-1 d, e and f.
\item We calculate present-day atmosphere abundances and outgassing rates of H$_2$, H$_2$O, CO, and CO$_2$.
\item Oxygen fugacity has a significant effect on the present-day abundances of H$_2$ and CO$_2$ on all three planets. 

\end{keypoints}

\begin{abstract}
The TRAPPIST-1 planets have become prime targets for studying the atmospheric and geophysical properties of planets around M-dwarf stars. To effectively identify their atmospheric composition, we first must understand their geological evolution. For this study, we focus on enhancing an existing atmosphere-interior exchange model by incorporating additional geological processes relevant to rocky planets. We have extended the model to include the carbon cycle, which enables the model to track four key gas species - CO$_2$, CO, H$_2$O, and H$_2$ - across four planetary reservoirs: the mantle, plate, ocean, and atmosphere. Major features added include surface temperature calculations which are crucial for the carbon cycle, oxygen fugacity as a planetary interior parameter in the model, and oxidation reactions and diffusion-limited escape calculations to the atmosphere portion of the model. We successfully validated the model for Earth and applied this model to study the effect of oxygen fugacity and initial water abundance on TRAPPIST-1 d, e and f. Our results for present-day abundances show that oxygen fugacity significantly affects the partial pressures of H$_2$ and CO$_2$ for all three planets with minor effects for CO on two of the planets. We also found that H$_2$ is strongly dependent on water mass fraction (WMF). The addition of atmospheric processes produced a significant difference in the H$_2$ and CO abundances at present-day. These results highlight the importance of considering interior parameters to be able to further constrain the geological evolution of these planets and effectively put atmosphere observations into context.
\end{abstract}

\section*{Plain Language Summary}
The TRAPPIST-1 system provides a natural laboratory for studying exoplanet habitability. Because these planets orbit stars that are not Sun-like, we are interested in learning whether these planets have atmospheres. We are also interested in learning what kind of geological processes shaped their atmospheres over time. For this study, we focus on the TRAPPIST-1 system, specifically planets d, e and f. We use a model that can simulate geological processes through tectonic evolution like that of Earth. We study how the interior, the surface, and the atmosphere of the planet interact with each other through two geochemical cycles - the deep-water cycle and the carbon cycle. With our improved model, we track various gases including H$_2$, H$_2$O, CO, and CO$_2$ to understand the evolution of the planets' atmospheres given different interior parameters, such as oxygen fugacity and initial water abundance. We find that oxygen fugacity strongly influences H$_2$ and CO$_2$ abundances in the atmosphere, while water abundance strongly impacts the H$_2$ abundance. These results highlight the importance of considering various parameters and geological processes to understand their context when interpreting atmosphere observations obtained with telescopes.

\section{Introduction}

Thousands of exoplanets have been discovered in the last several decades, increasing the need to characterize these exoplanets. Most stars found in our solar neighborhood are M dwarfs, which are cooler and redder stars than our Sun \cite{reyle_gaia}. Planets around M dwarfs are favorable for being detected through the transit technique - when the planet passes in front of their star and blocks part of the light that we receive from the star. Having a smaller star will produce deep transits for any given planet compared to a Sun-like star. Temperate planets around M-dwarfs also have a more favorable geometric probability of transit than temperate planets around Sun-like stars \cite{charbonneau2007, nutzman2008}.  This, coupled with the estimate that M dwarfs seem to have around $2.5 \pm 0.2$ planets with radii $1-4 R_{\oplus}$ per star \cite{dressing_occurrence}, makes these exoplanets prime targets for in-depth studies to further characterize them. The NASA Transiting Exoplanet Survey Satellite (TESS) mission has detected several of these types of planets \cite{ricker_tess} and combined with the capabilities of JWST, studies are underway to detect and characterize the atmospheres of these exoplanets around smaller stars. The TRAPPIST-1 system, which has seven small planets orbiting a M-dwarf star, has become one of the most promising planetary systems to study in the context of habitability \cite{gillon2017}. 

As demonstrated by the extensive literature on the TRAPPIST-1 system \cite{agol_2021, barth2021magma, krissansen-totton2022, boldog2024, van2024airy, gialluca2024implications, krissansen-totton2024erosion, thomas2025, rice2025uncertainties}, these planets have become the prime targets for habitability studies because there are multiple planets that are relatively easily observable. Not only do we want to be able to detect their atmospheres if they have any, but we also want to say something about their formation and evolution. Therefore, in this study, we focus on two of the planets in the conservative habitable zone of TRAPPIST-1 \cite{kopparapu2014} and one of the planets in the optimistic habitable zone (HZ). Through this study, we aim to explore different geologic and atmospheric evolution paths that these planets could have taken in their lifetimes and study how much their evolution has impacted the atmospheric composition that we would expect to see with telescopes today. 

JWST has pushed the boundaries of exoplanet science and has started to show that it can probe smaller, rocky planets \cite{greene2023,zieba2023,lustig-yager2023,moran2023,may2023,kirk2024,bennett2025}. While JWST is beginning to probe these planets, this is just the beginning of studying habitability elsewhere. To fully understand these exoplanets, we need to look inward to understand their geological evolution. Just as Earth's habitability has been influenced by its geological evolution over time, we expect exoplanets to be influenced in a similar manner. Therefore, we aim to explore how geological processes can influence atmospheric composition on such worlds.

We begin to explore the influence of geological processes on atmospheric processes by specifically studying how outgassing, a process through which volatiles move between the atmosphere and the interior of a planet, directly impacts the atmospheric composition of the planets in the TRAPPIST-1 system. Outgassing can occur as a result of geological cycling, through which volatiles are transported between the interior, surface, and atmosphere of a planet \cite{karato2015water,hirschmann2018comparative,sleep2001carbon}. Two of the cycles we focus on are the carbon cycle and the deep water cycle. We emphasize the importance of including these processes when studying the atmospheres of rocky exoplanets as they can have a role in setting atmospheric composition and climate of such planets.

The deep water cycle is a process through which water is removed from the atmosphere, deposited onto the surface of a planet and subducted into the mantle of Earth in the form of hydrated minerals in oceanic crust. This water can then be expelled from the mantle of a planet onto the surface and atmosphere through outgassing. This process is facilitated through plate tectonics on Earth, and has a significant role in cycling water in our planet \cite{hirschmann2006, schaefer2015persistence}, which has implications for long-term habitability.  The cycle has been modeled before \cite{mcgovernshubert1989, bounama2001, crowley2011, sandu_2011,schaefer2015persistence} through parameterized convection models and we build upon those for our study. Similarly, the carbon cycle \cite{walker1981} functions in the same way as the deep-water cycle in terms of how volatiles move throughout the planet. Carbon is removed from the atmosphere, interacts chemically with rocks on the surface and can be deposited at the bottom of the ocean on top of the oceanic plate, which is then subducted into the mantle of Earth. This carbon can then be outgassed through various forms of volcanism and mid-ocean ridges. This cycle is important on Earth because it acts as a stabilizer of the planet's temperature over long geologic timescales through a negative feedback cycle \cite{walker1981}. It has also previously been modeled for rocky planets \cite{foley2015, joshkt2017} and we aim to incorporate it for this study. We use the same framework for these cycles on Earth for this study on the TRAPPIST-1 planets. 

Using the combined deep water cycle and the carbon cycle will enable us to study how geological cycling can impact the atmospheric composition and its evolution over the planet's lifetime. But, as many other geological processes on Earth, they depend on an array of parameters. Two key parameters we aim to explore in this study are the oxygen fugacity and the water abundance of the planet. Oxygen fugacity (fO$_2$) is a way to measure oxidation of the rocks \cite{doyle2019}, particularly in the mantle, and can be used to determine how reducing or oxidizing a system is \cite{guimond2023}. Oxygen fugacity is intrinsically linked to the abundances of different iron (Fe) species. The mantle oxygen fugacity can be set by the different iron species of Fe$^{2+}$ (FeO) and Fe$^{3+}$ (Fe$_{2}$O$_{3}$). This is directly proportional to the ratios of partial pressures of gas species that are related to each other by the addition of oxygen to one of the species (e.g. H$_2$/H$_2$O, CO/CO$_2$) \cite{schaefer2017}. This relationship implies that, depending on the oxygen fugacity, the planet may exhibit a preference to outgas specific molecules over others. For example, having higher oxygen fugacities may lead to more CO$_{2}$ and H$_{2}$O than lower oxygen fugacities, which are more likely to outgas CO and H$_{2}$ \cite{kasting1993, guimond2023, schaefer2017}. Similarly, another key parameter that could potentially affect the outgassing of volatiles is the initial volatile abundances. The initial water inventory could have an effect on how it influences material properties such as phase changes, rheology, among others \cite{hirth1996,karato1990,karato1993}, which can affect the viscosity and therefore the material flow of the mantle. This could produce a change in the geological cycling of volatiles in a planet facilitated by plate tectonics, which would then have implications for the atmospheric composition of the planet and how it could vary over its lifetime.  

Ultimately, the goal is to study the temporal effects of geological cycling on the outgassing and the atmospheric composition on the TRAPPIST-1 planets. While there are similar models to the one used in this study \cite{krissansen-totton2022,krissansen-totton2024erosion,thomas2025}, we specifically want to test the effect of crucial interior parameters such as oxygen fugacity (fO$_2$) and initial volatile abundance for the TRAPPIST-1 planets on the resulting atmospheres. We also aim to provide estimated outgassing rates and partial pressures of the various gases depending on these two parameters at present day, providing a range of plausible abundances that could be observed with telescopes such as JWST. With this range, we could determine detection limits moving forward to observe these differences in transmission and emission spectra and begin to disentangle atmosphere and interior properties to provide much needed geological context of these exoplanets. 

In this paper, we further develop a deep volatile cycle model that computes the coupled thermal and volatile evolution of a rocky planet, and then apply the model to the TRAPPIST-1 planets. In Section \ref{sec:methods} we describe the outgassing model, which includes the deep-water cycle model, the carbon cycle model, surface temperature calculations, oxygen fugacity and outgassing, and the initial parameters used for the TRAPPIST-1 planets in our model. In Section \ref{sec:results} we discuss the results for these models for TRAPPIST-1 d, e and f, which include the nominal case for these planets, the effect of oxygen fugacity, and the effect of water abundance on the evolution of these planets. We also include results for the outgassing rates for these planets. In Section \ref{sec:discussion}, we compare our results to other models and discuss the implications of our results for the TRAPPIST-1 planets.

\section{Methods} \label{sec:methods}

\begin{figure}[t]
    \centering
%    \hspace*{-8mm}
    \includegraphics[width=\textwidth]{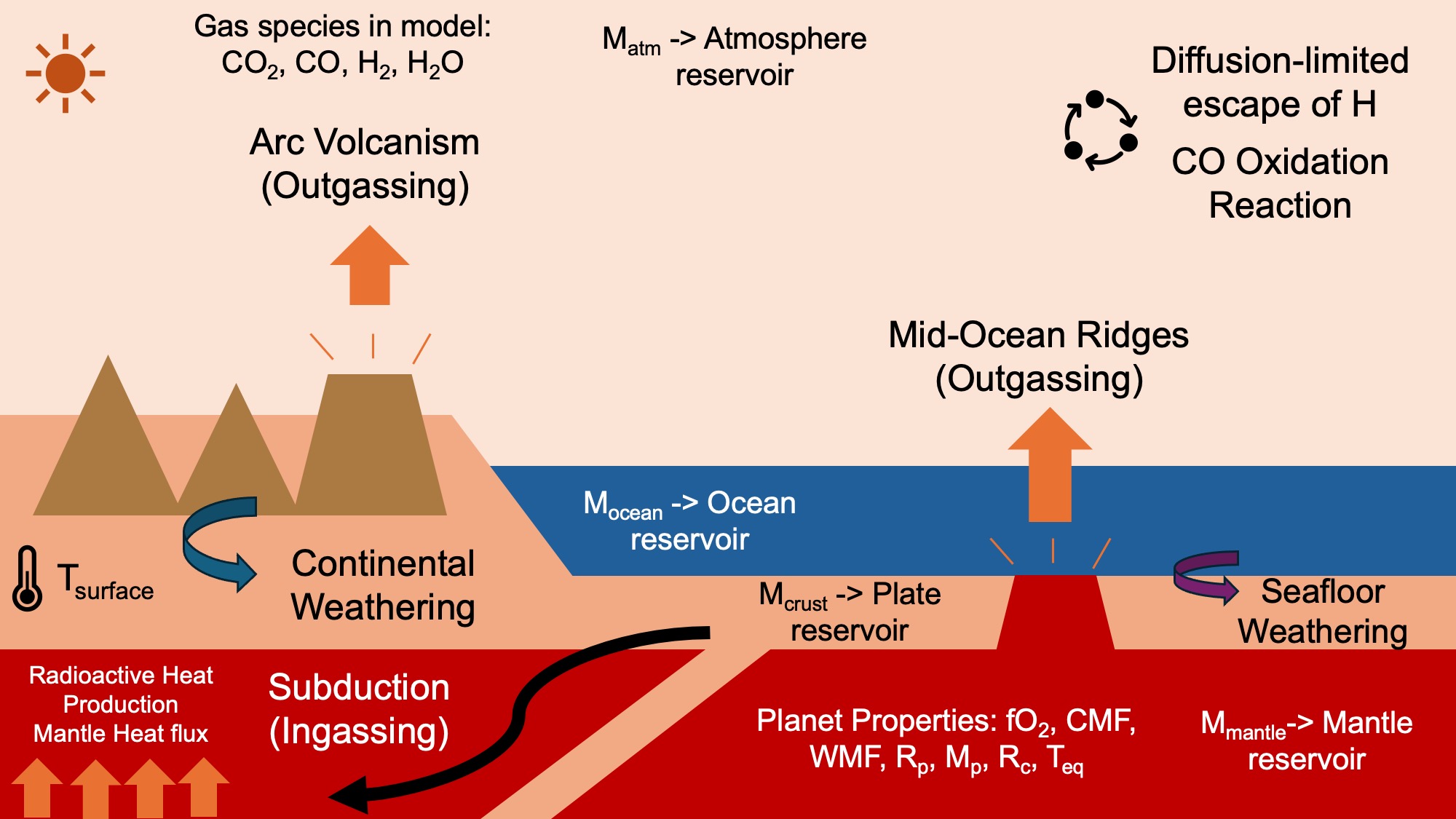}
    \caption{Overview of the atmosphere-interior exchange model we use for this study including the various geological processes that we simulate and the various parameters we consider in this model. The carbon cycle is the major overarching process added to this model, which is a negative feedback cycle that regulates the temperature of the planet over geologic timescales, which is crucial for studying habitability.}
    \label{fig:model-overview}
\end{figure}

Rocky exoplanet atmospheres can be a result of a combination of processes, including impacts during planet formation, as well as outgassing over time \cite{lichtenberg2023}. In this study, we are particularly interested in studying the role geological outgassing plays in the evolution of planets and their atmospheres. Using the deep water cycle model from \citeA{schaefer2015persistence} outlined in section \ref{sec:deepwatercycle}, we added the carbon cycle as described in section \ref{sec:carboncycle}, which allows us to track these four gas species - CO$_2$, CO, H$_2$, H$_2$O. We added two new atmospheric processes to the model, diffusion-limited hydrogen escape and a CO oxidation reaction, to include the interactions of gas species within the atmosphere in section \ref{sec:deepwatercycle}. We also discuss in section \ref{sec:Tsurf} that the carbon cycle regulates the surface temperature of a planet over long geological timescales and therefore added it to our model. Subsequently, we refer to the added feature of setting the oxygen fugacity within the mantle in section \ref{sec:fO2} as a way to test and explore the effect of this parameter on the cycling of volatiles on the planet. Lastly, in section \ref{sec:trappist1} we highlight the initial parameters used for the modeling of the TRAPPIST-1 planets. These processes add relevant considerations to the planetary evolution model that we can then use to study exoplanets, including the TRAPPIST-1 planets. This helps us explore the possible evolution paths that these planets could have taken during their lifetimes, more specifically their later stage evolution. In Figure \ref{fig:model-overview}, we show the overview of our atmosphere-interior exchange model that we use for this study with all processes and relevant parameters included.

\subsection{Deep-Water Cycle}\label{sec:deepwatercycle}

The deep-water cycle model in \citeA{schaefer2015persistence} explored the distribution of water in super-Earths experiencing an active plate tectonic regime. Although that model provided insight into the variability of the water distribution between the surface and the mantle of exoplanets, it did not account for other relevant gas species for habitability such as different carbon species (CO$_{2}$, CO) or other hydrogen-bearing species. A key factor called oxygen fugacity ($fO_{2}$) has been shown in various studies to influence the abundances of certain gas species that are outgassed over a planet's lifetime \cite{Schaefer2017redox,ortenzi2020mantle,gaillard2014,hirschmann2012,holloway1992high}. We therefore modified the model set forth in \citeA{schaefer2015persistence} in order to account for oxygen fugacity in the geological evolution of a planet and to track the abundances of additional gas species.

\afterpage{
\clearpage
 \LTcapwidth=\textwidth
 \begin{longtable}{l c c}
 \caption{Table of variables used within the deep water and carbon cycle models. The equation where the variable is first used is given in the right-hand column.} \\
 \hline
  Symbol & Definition and Units & Equation \\
 \hline
 $M^{H_{2}O}_{man}$ & Water abundance in the mantle (kg) & \ref{eq:m_water_mantle}\\
 $M^{H_{2}O}_{atm}$ & Water abundance in the atmosphere (kg) & \ref{eq:m_water_ocean}\\
  $M^{H_{2}O}_{oc}$ & Water abundance in the ocean (kg) & \ref{eq:m_water_ocean}\\
 $M^{H_{2}}_{atm}$ & Hydrogen abundance in the atmosphere (kg) & \ref{eq:m_hydrogen}\\
 $M^{H_{2}}_{oc}$ & Hydrogen abundance in the ocean (kg) & \ref{eq:m_hydrogen}\\
 $M^{CO_{2}}_{crust}$ & Carbon dioxide (CO$_{2}$) abundance in the crust (kg) & \ref{eq:plate_co2}\\
 $M^{CO_{2}}_{man}$ & Carbon dioxide (CO$_{2}$) abundance in the mantle (kg) & \ref{eq:mantle_co2}\\
 $M^{CO_{2}}_{atm}$ & Carbon dioxide (CO$_{2}$) abundance in the atmosphere (kg) & \ref{eq:atm_co2}\\
 $M^{CO_{2}}_{oc}$ & Carbon dioxide (CO$_{2}$) abundance in the ocean (kg) & \ref{eq:atm_co2}\\ 
 $M^{CO}_{atm}$ & Carbon monoxide (CO) abundance in the atmosphere (kg) & \ref{eq:m_co}\\
 $M^{CO}_{oc}$ & Carbon monoxide (CO) abundance in the ocean (kg) & \ref{eq:m_co}\\
 $r^{H_{2}O}_{ingas}$ & Ingassing rate of water (H$_{2}$O) (kg/m$^{2}$) & \ref{eq:m_water_mantle}, \ref{eq:m_water_ocean}\\
 $r^{H_{2}O}_{outgas}$ & Outgassing rate of water (H$_{2}$O) (kg/m$^{2}$) & \ref{eq:m_water_mantle}, \ref{eq:m_water_ocean}\\
 $r^{H_{2}}_{outgas}$ & Outgassing rate of hydrogen (H$_{2}$) (kg/m$^{2}$) & \ref{eq:m_water_mantle}, \ref{eq:m_hydrogen} \\
 $r^{H_{2}}_{ingas}$ & Ingassing rate of hydrogen (H$_{2}$) (kg/m$^{2}$) & \ref{eq:m_water_ocean} \\
 $r_{CO-ox}$ & CO-oxidation reaction rate (m$^3$ molecules$^{-1}$ sec$^{-1}$) & \ref{eq:r_photochem} \\
 $r_{CO-ox}^{CO}$ & Rate of CO lost due to CO oxidation reaction (m$^3$ molecules$^{-1}$ sec$^{-1}$) & \ref{eq:r_photochem}, \ref{eq:m_co} \\
 $r^{H_2O}_{CO-ox}$ & Rate of water (H$_{2}$O) lost due to CO oxidation (kg/m$^{2}$) & \ref{eq:m_water_ocean} \\
 $r^{H_2}_{CO-ox}$ & Rate of hydrogen (H$_{2}$) gained due to CO oxidation (kg/m$^{2}$) & \ref{eq:m_hydrogen}, \ref{eq:m_water_ocean} \\
 $r^{CO_{2}}_{CO-ox}$ & Rate of carbon dioxide (CO$_{2}$) gained due to CO oxidation (kg/m$^{2}$) & \ref{eq:m_water_ocean}, \ref{eq:atm_co2} \\
 $r^{CO_{2}}_{ingas}$ & Ingassing rate of CO$_{2}$ (kg/m$^{2}$) & \ref{eq:mantle_co2}, \ref{eq:plate_co2} \\
 $r_{weather}^{CO_2}$ & Continental weathering rate (kg/m$^{2}$) & \ref{eq:plate_co2}, \ref{eq:weathering}, \ref{eq:atm_co2} \\
 $r_{sfw}^{CO_2}$ & Seafloor weathering rate of CO$_2$ (kg/m$^{2}$) & \ref{eq:plate_co2}, \ref{eq:sfw}, \ref{eq:atm_co2} \\
 $r_{arc}^{CO_2}$ & Rate for arc volcanism of CO$_2$ (kg/m$^{2}$) & \ref{eq:atm_co2} \\
 $r_{outgas}^{CO_{2}}$ & Outgassing rate of carbon dioxide (CO$_{2}$) (kg/m$^{2}$) & \ref{eq:mantle_co2}, \ref{eq:atm_co2} \\
 $r_{outgas}^{CO}$ & Outgassing rate of carbon monoxide (CO) (kg/m$^{2}$) & \ref{eq:mantle_co2}, \ref{eq:m_co}, \ref{eq:rmorco} \\
  $r_{escape}^{H_2}$ & Rate of diffusion-limited escape (kg/m$^{2}$) & \ref{eq:m_hydrogen}, \ref{eq:escape} \\
 $k_{rate}$ & Second order rate of photochemical reaction (m$^{3}$ molecules$^{-1}$ sec$^{-1}$) & \ref{eq:r_photochem}, \ref{eq:krate} \\
 $[CO]$ & Number density of carbon monoxide (CO) (molecules/m$^{2}$) & \ref{eq:r_photochem}, \ref{eq:co}\\
 $[OH]$ & Number density of hydroxyl (OH) (molecules/m$^{2}$) & \ref{eq:r_photochem}, \ref{eq:oh}\\
 $f_{H}$ & Total atomic fraction of H$_{2}$ & \ref{eq:escape} \\
 $H_{i}$ & Atmospheric scale height (meters) & \ref{eq:scaleheight} \\
 $H_{N_{2}}$ & Scale height of nitrogen (N$_{2}$) in the atmosphere (meters) & \ref{eq:escape} \\
 $H_{H_{2}}$ & Scale height of hydrogen (H$_{2}$) in the atmosphere (meters) & \ref{eq:escape} \\
 $m_{i}$ & Molar weight of gas species i & \ref{eq:scaleheight} \\
 $P_{H_{2}}$ & Partial pressure of hydrogen (H$_{2}$) in the atmosphere (bars) & \ref{eq:Ph2} \\
 $P_{sat}$ & Saturated partial pressure of water (H$_{2}$O) in the atmosphere (bars) & \ref{eq:oh}, \ref{eq:Psat} \\
 $P_{CO}$ & Partial pressure of carbon monoxide (CO) in the atmosphere (bars) & \ref{eq:co}, \ref{eq:Pco} \\
 $P_{CO_{2}}$ & Partial pressure of carbon dioxide (CO$_{2}$) in the atmosphere (bars) & \ref{eq:Pco2}, \ref{eq:sfw}, \ref{eq:weathering}, \ref{eq:Tsurf_foley} \\
 $K_{I}$ & Equilibrium Constant for C-CO$_2$ reaction (eq.\ref{eq:CCO2rxn}) & \ref{eq:Xmeltcarbonate}, \ref{eq:KI} \\
 $K_{II}$ & Equilibrium Constant for CO$_2$-CO$^{2-}_{3}$ reaction (eq.\ref{eq:CO3rxn}) & \ref{eq:Xmeltcarbonate},\ref{eq:KII} \\
 $fO_{2}$ & Oxygen fugacity (bars) & \ref{eq:Xmeltcarbonate}, \ref{eq:k1}, \ref{eq:k2}, \ref{eq:oxygfug} \\
 $X^{melt}_{CO^{-2}_{3}}$ & Mass fraction of carbonate in the melt & \ref{eq:Xmeltcarbonate}, \ref{eq:Xmeltco2} \\
 $X^{melt}_{CO_{2}}$ & Mass fraction of carbon dioxide (CO$_{2}$) in the melt & \ref{eq:Xmeltco2} \\
 $M_{ocean}$ & Fraction of total initial water in the ocean (kg) & \ref{eq:Ph2}, \ref{eq:Pco}, \ref{eq:Pco2} \\
 $M_{CO_2}$ & Mass of carbon dioxide present (kg) & \ref{eq:Pco2} \\
 $M_{CO}$ & Mass of carbon monoxide (CO) present (kg) & \ref{eq:Pco} \\
 $M_{H_2}$ & Mass of hydrogen (H$_2$) present (kg) & \ref{eq:Ph2} \\
$R_{p}$ & Radius of the planet (km) & \ref{eq:Ph2}, \ref{eq:Pco}, \ref{eq:Pco2} \\
 $g$ & Gravity of planet (m/s$^{2}$) & \ref{eq:scaleheight}, \ref{eq:Ph2}, \ref{eq:Pco}, \ref{eq:Pco2} \\
 $T_{surface}$ & Surface temperature of planet (K) & \ref{eq:Tsurf_foley} \\
$T_{eq}$ & Equilibrium temperature of planet (K) & \ref{eq:teq}, \ref{eq:co}, \ref{eq:oh}, \ref{eq:krate}, \ref{eq:scaleheight} \\
$T_{p}$ & Mantle potential temperature (K) & \ref{eq:krate} \\
$v$ & Spreading velocity (m/s) & \ref{eq:sfw} \\
$r_{w_{s}}$ & Supply limit to weathering & \ref{eq:weathering} \\
r$^{i}_{outgas}$ & Outgassing flux for volatile i & \ref{eq:rmor} \\
$\chi^{i}_{d}$ & Degassing efficiency of volatile i & \ref{eq:rmor} \\
$\rho_{m}$ & Density of the mantle & \ref{eq:rmor} \\
F$_{melt}$ & Average melt fraction & \ref{eq:rmor} \\
$\chi^{i}_{melt}$ & Average volatile fraction in the melt & \ref{eq:rmor} \\
D$_{melt}$ & Thickness of the melt layer & \ref{eq:rmor} \\
S & Areal spreading rate  & \ref{eq:rmor} \\
K$_{30}$ & Equilibrium constant of reactions & \ref{eq:k1} \\
K$_{31}$ & Equilibrium constant of reactions & \ref{eq:k2} \\ 
$X^{gas}_{H_{2}}$ & Mole fraction of species H$_{2}$ in the gas phase & \ref{eq:k1}\\
$X^{gas}_{H_{2}O}$ & Mole fraction of species H$_{2}$O in the gas phase & \ref{eq:k1}\\
$X^{gas}_{CO}$ & Mole fraction of species CO in the gas phase & \ref{eq:k2}\\
$X^{gas}_{CO_{2}}$ & Mole fraction of species CO$_{2}$ in the gas phase & \ref{eq:k2}\\
X$_{H_{2}}$ & Relative abundance of species H$_{2}$ & \ref{eq:k1}\\
X$_{H_{2}O}$ & Relative abundance of species H$_{2}$O & \ref{eq:k1}\\
X$_{CO}$ & Relative abundance of species CO & \ref{eq:k2}\\
X$_{CO_{2}}$ & Relative abundance of species CO$_{2}$ & \ref{eq:k2}\\

 \hline
 \label{tab:variables}
\end{longtable}
}

The model developed in \citeA{schaefer2015persistence},  which we have adopted in this study, is a single layer convection model, which couples the thermal evolution of a planet's mantle with a volatile evolution model. The thermal evolution model calculates the rate of change of mantle temperature due to the balance between heat loss from the surface and internal heat production from radioactive decay of long-lived isotopes. The exchange of water between the surface and interior is dependent on the rate of tectonic plate motion, which is determined by mantle convection and the temperature-dependent mantle viscosity. Water is outgassed from volcanic centers via melting and outgassing and is returned to the mantle through subducting oceanic plates, which trap water in minerals and sediments. For details on the implementation of the thermal evolution model and the deep-water cycle model, we refer readers to \citeA{schaefer2015persistence}. Here we discuss the aspects of the deep-water cycle that we modified for the purposes of this study.

As in \citeA{schaefer2015persistence}, we calculate the rate of change of the water abundance in the mantle by accounting for the ingassing ($r_{ingas}^{H_{2}O}$) and outgassing ($r_{outgas}^{H_{2}O}$) rates of water. We are particularly interested in adding and tracking more gas species to improve our understanding of how outgassing can affect the atmospheric composition of a planet. Therefore, we additionally track the amount of H$_{2}$ that is outgassed ($r_{outgas}^{H_{2}}$). We assume that all of the hydrogen in the solid mantle is dissolved as H$_{2}$O, so outgassing of H$_{2}$ acts as a sink for mantle water, as shown in the following equation: 

\begin{equation}
\frac{dM^{H_{2}O}_{man}}{dt} = r^{H_{2}O}_{ingas} - r^{H_{2}O}_{outgas} - r^{H_2}_{outgas}
\label{eq:m_water_mantle}
\end{equation}

\citeA{schaefer2015persistence} did not explicitly track the water in the plate or atmosphere because it can be calculated by mass-balance in a closed system. However, in our updated model, we adopt an open system that allows atmospheric escape, so we have added an equation to explicitly track the evolution of the surface reservoir of water (plate + ocean + atmosphere) in our model as shown in the following equation:

\begin{equation}
\frac{dM^{H_{2}O}_{atm}+M^{H_{2}O}_{oc}}{dt} =  r^{H_{2}O}_{outgas} - r^{H_{2}O}_{ingas} - r^{H_{2}O}_{CO-oxidation}
\label{eq:m_water_ocean}
\end{equation}
where, in addition to the ingassing sink and outgassing source, we have added a simple but influential chemical reaction to account for how the atmospheric chemistry can modify the outgassed atmospheric composition over long timescales. The CO-oxidation reaction is given in eq. \eqref{eq:chemreac}. In eq. \eqref{eq:m_water_ocean}, $r_{CO-oxidation}^{H_{2}O}$ is the destruction rate of atmospheric water vapor due to the CO-oxidation reaction given by:

\begin{equation}
CO + OH \rightarrow CO_{2} + H
\label{eq:chemreac}
\end{equation}

While the reaction includes OH, we assume that water vapor (H$_{2}$O) is efficiently converted into OH in the upper atmosphere via photolysis in order to maintain the simplicity of our model. The number of atmospheric chemistry reactions included could be expanded in future work, but the selected reaction is the fastest, and therefore likely dominant, reaction for oxidizing CO gas. The reaction rate ($r_{CO-oxidation}$) is given by multiplying the second-order rate coefficient (in m$^{3}$ molecules$^{-1}$ sec$^{-1}$) by the number densities of carbon monoxide (CO) and hydroxyl (OH) (in molecules m$^{-2}$).

\begin{equation}
r_{CO-oxidation} = k_{rate} [CO] [OH]
\label{eq:r_photochem}
\end{equation}
where $k_{rate}$ is given by the equation from \citeA{baulch_1992}:

\begin{equation}
k_{rate} = A \left( \frac{T_{eq}}{298}\right) ^n \exp{\left( \frac{-E_{a1}}{R T_{eq}}\right) }
\label{eq:krate}
\end{equation}

To calculate the equilibrium temperature for each planet, we follow the equation:

\begin{equation}
T_{eq} = T_{eff}^{*} (1-a_{p})^{1/4} (\frac{R_{star}}{2a})^{1/2}
\label{eq:teq}
\end{equation}

where $T_{eff}^{*}$ is the effective temperature of the star, $a_p$ is the albedo, $R_{star}$ is the radius of the star and $a$ is the semi-major axis of each planet as given in Table \ref{tab:constants}. We choose an albedo of 0.3, similar to Earth, as it does not produce a significant difference in the surface temperature parameterization for this study. However, studies in the effects of albedo have been made \cite{rushby2020} and could be explored further in future studies.

\begin{table}
\caption{Table of physical constants and their values}
\begin{tabular}[h]{l c c c}
\hline
 Symbol & Definition and Units & Value & Equation \\
\hline
$P_{sat0}$ & Reference saturated water vapor pressure (Pa) & 610 & \ref{eq:Psat}\\
$P^{*}_{CO_{2}}$ & Present day atmospheric CO$_2$ (Pa) & 33 & \ref{eq:sfw}, \ref{eq:weathering}, \ref{eq:Tsurf_foley}\\
$T_{sat0}$ & Reference temperature for water vapor (K) & 273 & \ref{eq:Psat} \\
$\bar{m}_{w}$& Molar weight of water (g mole$^{-1}$)&  18.015 & \ref{eq:Psat}\\
$L_{w}$ & Latent heat of water (J g$^{-1}$) & 2469 & \ref{eq:Psat} \\
$r^{*}_{sfw}$ & Present day seafloor weathering flux (mol Ma$^{-1}$) & $1.75 \times 10^{18}$ & \ref{eq:sfw} \\
$v^*$ & Present day plate speed (cm yr$^{-1}$) & 5 & \ref{eq:sfw} \\
$\alpha$ & P$_{CO_{2}}$ Exponent for seafloor weathering & 0.25 & \ref{eq:sfw} \\
$R$& Ideal gas constant (J mole$^{-1}$ K$^{-1}$) & 8.314 &  \ref{eq:Psat}\\
$A$ & Kinetic rate coefficient (m$^{3}$ molecules$^{-1}$ sec$^{-1}$) & $5.4 \times 10^{-14}$ & \ref{eq:krate}\\
$E_{a1}$ & Activation energy (J mole$^{-1})$ & -2,079 & \ref{eq:krate} \\
$n$ & Kinetic rate exponent & 1.50 & \ref{eq:krate}\\
$N_A$ & Avogadro's number (molecules mole$^{-1}$) & $6.022 \times 10^{23}$ & \ref{eq:co}, \ref{eq:oh}, \ref{eq:Pco} \\
$b_H$ & Binary diffusion coefficient of H$_2$ through N$_2$& $18.8 \times 10^{16} T^{0.82}$ & \ref{eq:escape}\\
 &  (molecules cm$^{-1}$ sec$^{-1}$)& &\\
$k_B$ & Boltzmann's constant (J K$^{-1}$) & $1.38 \times 10^{-23}$ & \ref{eq:scaleheight}\\
$k_{H_2}$ & Solubility of H$_2$ in water (mole L$^{-1}$ bar$^{-1}$) & $7.8 \times 10^{-4}$ & \ref{eq:Ph2}\\
$k_{CO}$ & Solubility of CO in water (mole L$^{-1}$ bar$^{-1}$) & $1.0 \times 10^{-3}$ & \ref{eq:Pco} \\
$k_{CO_2}$ & Solubility of CO$_2$ in water (mole L$^{-1}$ bar$^{-1}$) & $2.3 \times 10^{-1}$ & \ref{eq:Pco2}\\
$f$ & Fraction of CO$_2$ in plate that is not subducted & 0.5 & \ref{eq:mantle_co2}\\
$r^{*}_{w}$ & Present day weathering flux (mol Ma$^{-1}$) & $12 \times 10^{18}$ & \ref{eq:weathering} \\
f$_{land}$ & Fraction of exposed land & 0.3 & \ref{eq:weathering}, \ref{eq:supplylimit} \\
f$^{*}_{land}$ & Present day land fraction & 0.3 & \ref{eq:weathering} \\
$\beta$ & P$_{CO_{2}}$ Exponent for silicate weathering & 0.55 & \ref{eq:weathering} \\
a & P$_{sat}$ Exponent for silicate weathering & 0.3 & \ref{eq:weathering} \\
E$_{a2}$ & Activation energy for silicate weathering (kJ mol$^{-1}$) & 42 & \ref{eq:weathering} \\
R$_{g}$ & Gas constant (J K$^{-1}$ mol$^{-1}$) & 8.314 & \ref{eq:co}, \ref{eq:oh}, \ref{eq:krate}, \ref{eq:Psat}, \ref{eq:weathering} \\
T$^{*}$ & Present day surface temperature (K) & 285 & \ref{eq:weathering}, \ref{eq:Tsurf_foley} \\
T$^{*}_{e}$ & Present day effective temperature (K) & 254 & \ref{eq:Tsurf_foley} \\
T$_{eff}^{*}$ & Effective temperature of TRAPPIST-1 (star) (K) & 2566 K & \ref{eq:teq} \\
A$_{Earth}$ & Surface area of the Earth (km $^{2}$) & 5.1 $\times 10^{14}$  & \ref{eq:supplylimit} \\
E$_{max}$ & Maximum erosion rate (mm yr$^{-1}$) & 10 & \ref{eq:supplylimit}  \\
$\chi_{cc}$ & Fraction of Mg, Ca, K, and Na in continental crust & 0.08 & \ref{eq:supplylimit} \\
$\rho_{r}$ & Density of regolith (kg m$^{-3}$) & 2500 & \ref{eq:supplylimit} \\
$\bar{m}_{cc}$ & Average molar mass of Mg, Ca, K, and Na (g mol$^{-1}$) & 32 & \ref{eq:supplylimit} \\
a$_p$ & Albedo for all planets & 0.3 & \ref{eq:teq}, \ref{eq:Tsurf_foley} \\
$\omega$ & OH concentration constraint $\times$ water& 10$^{-17}$ & \ref{eq:oh} \\
R$_{star}$ & Radius of the TRAPPIST-1 star (R$_{\odot}$) & 0.1192 & \ref{eq:teq}  \\
a$_{planet-d}$ & Semi-major axis of TRAPPIST-1 d (AU) & 2.227 $\times$ 10$^{-2}$ & \ref{eq:teq} \\
a$_{planet-e}$ & Semi-major axis of TRAPPIST-1 e (AU) & 2.925 $\times$ 10$^{-2}$ & \ref{eq:teq} \\
a$_{planet-f}$ & Semi-major axis of TRAPPIST-1 f (AU) & 3.849 $\times$ 10$^{-2}$ & \ref{eq:teq} \\
\hline
\label{tab:constants}
\end{tabular}
\end{table}

The number densities (in molecules m$^{-2}$) for CO and OH are given by the following equations: 

\begin{equation}
[CO] = \frac{P_{CO} N_{A}}{R T_{eq}}
\label{eq:co}
\end{equation}

\begin{equation}
[OH] = \frac{\omega P_{sat} N_{A}}{R T_{eq}}
\label{eq:oh}
\end{equation}

where we multiply the partial pressures ($P_{i}$) of each gas by Avogadro's number and divide by the ideal gas constant $R$ and the equilibrium temperature of the planet ($T_{eq}$). In the case of OH, we have added an additional parameter $\omega$, which is an empirically-determined scaling factor for the amount of OH produced in the upper atmosphere from photodissociation of water vapor, based on photochemical models of the TRAPPIST-1 planets from \citeA{wolf2025}. We also use the saturated water vapor partial pressure, which is given by \citeA{foley2015}:

\begin{equation}
P_{sat} = P_{sat0} \exp{\left(-\frac{\bar{m_{w}}L_{w}}{R} \left(\frac{1}{T}-\frac{1}{T_{sat0}}\right)\right)}
\label{eq:Psat}
\end{equation}
where $P_{sat0}$ is the reference saturated vapor pressure at the reference temperature $T_{sat0}$, $\bar{m}_{w}$ is the molar mass of water and $L_w$ is the latent heat of water. 

In addition to water, we track hydrogen in the atmosphere and ocean. To do so we account for hydrogen outgassing ($r_{outgas}^{H_{2}}$), the amount of hydrogen lost to diffusion-limited escape ($r^{H_2}_{escape}$), and the production of hydrogen through the CO-oxidation reaction ($r^{H_2}_{CO-oxidation}$). The evolution equation for hydrogen gas is then given by:

\begin{equation}
\frac{d(M^{H_{2}}_{atm}+M^{H_{2}}_{oc})}{dt} = r^{H_{2}}_{outgas} - r^{H_2}_{escape} + r^{H_2}_{CO-oxidation}
\label{eq:m_hydrogen}
\end{equation}

We do not track hydrogen in the mantle, since we assume all mantle hydrogen is in the form of water and is therefore fully accounted for by equation (\ref{eq:m_water_mantle}). The previous deep-water cycle model \cite{schaefer2015persistence} did not account for diffusion-limited escape, so we add a simple calculation based on \citeA{zahnle2008photochemical}. The escape flux is given by:  

\begin{equation}
r^{H_2}_{escape} = b_{H}f_{H}\left(\frac{1}{H_{N_{2}}}-\frac{1}{H_{H_{2}}}\right)
\label{eq:escape}
\end{equation}
where $b_{H}$ is the binary diffusion coefficient of $H_{2}$ through $N_{2}$, $f_{H}$ is the total atomic fraction of hydrogen ($=2f_{H_2}$), $H_{N_{2}}$ is the scale height of $N_{2}$ in the atmosphere, and $H_{H_{2}}$ is the scale height of $H_{2}$ in the atmosphere. The scale heights are calculated using the following equation:

\begin{equation}
H_{i} = \frac{k_B T_{eq}}{m_i g}
\label{eq:scaleheight}
\end{equation}

where $k_B$ is Boltzmann's constant, $T_{eq}$ is the equilibrium temperature, $m_i$ is the molar weight of the gas species $i$ and $g$ is the gravity of the planet. 

We additionally consider the partitioning of hydrogen H$_{2}$ gas between the atmosphere and the dissolved component in the ocean. We calculate the partial pressure of hydrogen using Henry's law, given by: 

\begin{equation}
P_{H_{2}} = \frac{M_{H_{2}}}{\left(k_{H_{2}}M_{ocean}+\frac{4 \pi R_{p}^{2}}{g}\right)}
\label{eq:Ph2}
\end{equation}
which is a function of the total hydrogen mass ($M_{H_{2}} = M^{H_{2}}_{atm}+M^{H_{2}}_{oc}$), solubility ($k_{H_{2}}$) (taken from \citeA{catling2017atmospheric}), mass of the ocean ($M_{ocean}$) and the radius and gravity of the planet. While the mass balance equations take into account ingassing and outgassing rates of the various gas species to track the total surface reservoirs, the specific partial pressures of each H-bearing gas in the atmosphere from these geochemical cycles are given by eq. \eqref{eq:Psat} and eq. \eqref{eq:Ph2}. 

Note that there is no ingassing flux for $H_2$ gas. This is because we assume that all hydrogen bound in rocks is in the form of H$_2$O (or OH). We will discuss how the outgassing flux of both H$_2$ and H$_2$O are calculated with respect to oxygen fugacity in Section \ref{sec:fO2}.

\subsection{Carbon Cycle} \label{sec:carboncycle}

Similar to the deep-water cycle, the carbon cycle is a geochemical process in which carbon is cycled through various layers of the planet, including the atmosphere, the surface and the mantle. For our model, we use the mass balance equations set forth by \citeA{foley2015} for tracking CO$_{2}$ in three main reservoirs: oceanic plate, mantle, and the combined atmosphere and ocean system. As in \citeA{foley2015}, we do not take into account the continental reservoir because it does not significantly affect the cycle feedback and the climate evolution of a planet. The most significant contribution of carbonates occurs on the seafloor and the carbon continental outgassing depends on the plate speed similar to that of the mantle outgassing \cite{foley2015, driscoll2013, sleep2001carbon}. In our modified version of this model, we also consider CO gas as well as CO$_{2}$ gas, but assume all mantle carbon is stored in the form of CO$_{2}$. The mantle carbon evolution equation is given by: 

\begin{equation}
\frac{dM^{CO_{2}}_{man}}{dt} = (1-f) r_{ingas}^{CO_{2}} - r_{outgas}^{CO_{2}} - r_{outgas}^{CO}
\label{eq:mantle_co2}
\end{equation}

where $r_{ingas}^{CO_{2}}$ is the ingassing rate, $r_{outgas}^{CO_{2}}$ is the outgassing rate for carbon dioxide ($CO_{2}$), $r_{outgas}^{CO}$ is the outgassing rate for carbon monoxide ($CO$), and $f$ is the fraction of subducted $CO_{2}$ that is immediately released through arc volcanism. We will discuss how the outgassing fluxes are determined in Section \ref{sec:fO2}. 

The evolution of the crustal reservoir includes sources from weathering reactions and loss to the mantle through subduction. It is described by the equation: 

\begin{equation}
\frac{dM^{CO_{2}}_{crust}}{dt} = \frac{r^{CO_2}_{weather}}{2} + r^{CO_2}_{sfw} - r^{CO_2}_{ingas}
\label{eq:plate_co2}
\end{equation}

where $r^{CO_2}_{weather}$ is the continental weathering rate and $r^{CO_2}_{sfw}$ is the seafloor weathering rate. We assume that the continental weathering flux is divided by two because it is assumed that half of the carbon drawn through weathering on land is re-released to the atmosphere when carbonates form on the ocean floor. It essentially represents a net flux of CO$_2$ from the different reservoirs through continental weathering \cite{foley2015}.  The seafloor weathering rate is taken from \citeA{foley2015} and is given by:

\begin{equation}
r^{CO_2}_{sfw} = r^{*}_{sfw} (\frac{v}{v^{*}}) (\frac{P_{CO_{2}}}{P^{*}_{CO_{2}}})^\alpha
\label{eq:sfw}
\end{equation}

where $r^{*}_{sfw}$ is the present day weathering flux, $v$ is the plate speed, $v^{*}$ is the present day plate speed, $P_{CO_2}^{*}$ is the present day atmospheric CO$_2$, $P_{CO_2}$ is the partial pressure of CO$_2$, and $\alpha$ is the $P_{CO_2}$ exponent for seafloor weathering. All present day values are Earth-based assumed values as stated in Table \ref{tab:constants}.

The continental weathering rate is also obtained from \citeA{foley2015}:

\begin{equation}
r^{CO_2}_{weather} = r_{w_{s}} \times \Bigg\{ 1 - exp \Bigg[ - \frac{r^*_w f_{land}}{r_{w_{s}} f^*_{land}} \Bigg(\frac{P_{CO_{2}}}{P^{*}_{CO_{2}}} \Bigg)^\beta \Bigg(\frac{P_{sat}}{P_{sat0}} \Bigg)^a \times exp \Bigg( \frac{E_{a2}}{R_g} \Bigg( \frac{1}{T^*} - \frac{1}{T} \Bigg) \Bigg) \Bigg] \Bigg\}
\label{eq:weathering}
\end{equation}

where $r^*_w$ is the present day weathering flux, $f^*_{land}$ is present day land fraction, $f_{land}$ is the fraction of exposed land, $\beta$ is the $P_{CO_2}$ exponent for silicate weathering and a is the $P_{sat}$ exponent for silicate weathering. E$_{a2}$ is the activation energy for silicate weathering, R$_g$ is the gas constant and $T^*$ is present day surface temperature. All other variables are described in previous equations and present-day values are Earth-based assumptions. $r_{w_{s}}$ is the supply limit to weathering given by:

\begin{equation}
r_{w_{s}} = \frac{A_{Earth} f_{land} E_{max} \chi_{cc} \rho_{r}}{\bar{m}_{cc}}
\label{eq:supplylimit}
\end{equation}

where $A_{Earth}$ is the surface area of Earth, $E_{max}$ is the maximum erosion rate, $\chi_{cc}$ is the fraction of Mg, Ca, K, and Na in continental crust, $\rho_{r}$ is the density of regolith and $\bar{m}_{cc}$ is the average molar mass of Mg, Ca, K, and Na.

The evolution of CO$_2$ in the combined atmosphere and ocean system includes sinks from weathering reactions and sources from outgassing and the oxidation of CO gas. It is given by the equation:

\begin{equation}
\frac{d(M^{CO_{2}}_{atm}+M^{CO_{2}}_{oc})}{dt} = r^{CO_{2}}_{arc} + r^{CO_{2}}_{outgas} - \frac{r^{CO_{2}}_{weather}}{2} - r^{CO_{2}}_{sfw} + r^{CO_{2}}_{CO-oxidation}
\label{eq:atm_co2}
\end{equation}
where $r^{CO_{2}}_{arc}$ is the rate for arc volcanism (given by $r^{CO_{2}}_{arc} = fr_{ingas}^{CO_{2}}$), and $r^{CO_{2}}_{CO-oxidation}$ is the rate of production of $CO_{2}$ gas from the oxidation reaction given by eq. \ref{eq:chemreac}.

As mentioned in Section \ref{sec:deepwatercycle}, oxygen fugacity is a key parameter to consider when studying the abundance of carbon species, especially stemming from geochemical cycling. In order to make the connection between their abundance and oxygen fugacity, we first consider the amount of carbonate in the melt, as set forth in \citeA{GROTT2011}. The concentration of carbonate in the melt $X^{melt}_{CO^{-2}_{3}}$ is given by:

\begin{equation}
x^{melt}_{CO^{-2}_{3}} = \frac{K_{II}K_{I}fO_{2}}{1+K_{II}K_{I}fO_{2}}
\label{eq:Xmeltcarbonate}
\end{equation}
where $K_{I}$ is the equilibrium constant from the reaction of $CO_{2}$ forming from graphite in a graphite-saturated system as seen in eq. \eqref{eq:CCO2rxn}. 

\begin{equation}
C(graphite) + O_{2} (fluid) = CO_{2} (fluid)
\label{eq:CCO2rxn}
\end{equation}

\begin{equation}
log_{10}K_{I} = 40.07639 - 2.53932 \times 10^{-2} T + 5.2709e-6 T^{2} + 0.0267 (P-1)/T  
\label{eq:KI}
\end{equation}

Similarly, $K_{II}$ is the equilibrium constant for the chemical reaction that produces carbonate ions ($CO^{2-}_{3}$) from $CO_{2}$ as seen in \eqref{eq:CO3rxn}. 

\begin{equation}
CO_{2}^{fluid} + O^{2-(melt)} = CO_{3}^{2-(melt)}
\label{eq:CO3rxn}
\end{equation}

\begin{equation}
log_{10}K_{II} = -6.24763 - 282.56/T - 0.119242(P-1000)/T
\label{eq:KII}
\end{equation}
These equilibrium constants are calculated as a function of pressure (in bars) and temperature (in K) using equations \ref{eq:KI} and \ref{eq:KII} from \citeA{holloway1992high}.

From eq. \eqref{eq:Xmeltcarbonate}, we can calculate the mole fraction of carbonate ion dissolved in the silicate melt with a given $K_{I}$, $K_{II}$ and an assumed $fO_{2}$. We can then  calculate the mass fraction of CO$_{2}$ in the melt that could be outgassed, by converting concentrations by mole into concentrations by mass using the following equation:

\begin{equation}
X^{melt}_{CO_{2}} = \frac{1.202656x^{melt}_{CO^{2-}_{3}}}{1+0.202656x^{melt}_{CO^{2-}_{3}}}
\label{eq:Xmeltco2}
\end{equation}

%Using the concentration of $CO_{2}$ in the melt from eq. \eqref{eq:Xmeltco2}, we can furthermore calculate how much of it is specifically carbon. This allows us to then go from carbon to how much $CO$ and $CO_{2}$ is being outgassed. More on this can be found in section 2.3.
We will discuss the calculation of the outgassing fluxes for CO$_2$ and CO in Section \ref{sec:fO2}.

We assume that CO is only found in the atmosphere and ocean, so we include only one equation to track its abundance. The evolution equation for CO includes a source from outgassing and a sink from the oxidation of CO to CO$_2$ (see eq. (\ref{eq:r_photochem})). The evolution equation is given by:

\begin{equation}
\frac{d(M^{CO}_{atm}+M^{CO}_{oc})}{dt} = r^{CO}_{outgas} - r^{CO}_{CO-oxidation}
\label{eq:m_co}
\end{equation}

The partitioning of CO between the atmosphere and ocean is governed by Henry's law and is given by:

\begin{equation}
P_{CO} = \frac{M_{CO}}{(k_{CO}M_{ocean}+\frac{4 \pi R_{p}^{2}}{g})}
\label{eq:Pco}
\end{equation}
where $k_{CO}$ is the solubility of CO gas in water, taken from \citeA{catling2017atmospheric}. We adopt a similar relationship for solubility of CO$_2$ in the ocean:

\begin{equation}
P_{CO_{2}} = \frac{M_{CO_{2}}}{\left(k_{CO_{2}}M_{ocean}+\frac{4 \pi R_{p}^{2}}{g}\right)}
\label{eq:Pco2}
\end{equation}
where we adopt the solubility of CO$_{2}$ in water from \citeA{foley2015}. 

In the following section, we discuss how we calculate the surface temperature of the planet, which depends on the partial pressure of CO$_2$ in the atmosphere. 

\subsection{Surface Temperature Calculations} \label{sec:Tsurf}

The carbon cycle regulates surface temperature over long timescales by modulating the pressure of CO$_2$ in the atmosphere. The rate of mantle convection, which controls the rates of both subduction and outgassing, depends on the temperature contrast between the surface and the upper mantle, so it is important to include surface temperature calculations. For calculating the surface temperature, we adopt the simple parameterization from \citeA{foley2015}, which is based on understanding of the forcing of climate on Earth by CO$_2$. The surface temperature is given by the equation: 

\begin{equation}
T_{surface} = T^{*} + 2 \left(T_{eq} - T_{e}^{*}\right) + 4.6 \left(\frac{P_{CO_{2}}}{P^{*}_{CO_{2}}}\right)^{0.346} - 4.6
\label{eq:Tsurf_foley}
\end{equation}

where $T^{*}$ is the present day surface temperature of Earth, $T_{e}$ is the effective temperature of our model planet, $T_{e}^{*}$ is  the present-day effective temperature of the Earth, and $P^{*}_{CO_{2}}$ is the current partial pressure of $CO_2$ in the modern Earth atmosphere. We acknowledge that this simple model neglects the effects of other sources of climate forcing, including the greenhouse effect of $H_2$ and of H$_2$O at high partial pressures. We also acknowledge the potential effect of atmospheric circulation on the surface temperature due to the TRAPPIST-1 planets likely synchronously rotating, which would have an effect on their climates \cite{wolf2017,turbet2018}. We defer the incorporation of an improved climate model to future work, since our focus here is not on the climate of the planets but on their overall atmospheric abundances. 

\subsection{Oxygen Fugacity and Outgassing}\label{sec:fO2}

 Oxygen fugacity has two important effects to consider related to outgassing: 1) it directly affects the distribution of elements between different species in the gas phase, and 2) it alters the solubility of carbon in silicate melts (the solubility of hydrogen as water is minimally affected by oxygen fugacity). Therefore, it is important to study how different oxygen fugacities can change the outgassing of various gas species in a system experiencing active plate tectonics and geochemical cycling like Earth. 

As described in Section \ref{sec:carboncycle}, oxygen fugacity is important for determining the amount of CO$_{2}$ in the melt (see equations \ref{eq:Xmeltcarbonate} - \ref{eq:Xmeltco2}).  As mentioned above, oxygen fugacity does not directly affect the solubility of water in the melt. Although the solubility of H$_2$ gas is dependent on $fO_2$, we neglect it here due to its relatively low abundance compared to dissolved water. Once we have determined the abundance of dissolved volatiles in the melt, we can determine the outgassing flux of the different gas species.  We first calculate the total outgassing fluxes of hydrogen and carbon using the equation: 

\begin{equation}
r^i_{outgas} = \chi^i_{d} \rho_{m} \langle F_{melt} \rangle \langle X^i_{melt} \rangle D_{melt} S 
\label{eq:rmor}
\end{equation}
where $\chi^i_{d}$ is the degassing efficiency of volatile $i$, $\rho_{m}$ is the density of the mantle, $\langle F_{melt} \rangle$ is the average melt fraction, $\langle X^i_{melt} \rangle$ is the average volatile fraction in the melt (where $i$ is either carbon or hydrogen), $D_{melt}$ is the thickness of the melt layer, and $S$ is the areal spreading rate, which is calculated from the thermal evolution model \cite{schaefer2015persistence}.

Using these total outgassing fluxes, we then convert to the resulting outgassing fluxes for H$_{2}$, H$_{2}$O, CO, and CO$_{2}$. We do this by considering the reactions:

\begin{equation}
H_{2}O \rightarrow H_{2} + \frac{1}{2} O_{2}
\label{eq:cr_k1}
\end{equation}

\begin{equation}
CO_{2} \rightarrow CO + \frac{1}{2} O_{2}
\label{eq:cr_k2}
\end{equation}

We use expressions for the equilibrium constants of these reactions from \citeA{Schaefer2017redox}, given by:

\begin{equation}
log_{10}K_{\ref{eq:cr_k1}} = \frac{-12794}{T_{p}} + 2.7768 = \frac{X^{gas}_{H_2} f^{1/2}_{O_2}}{X^{gas}_{H_2O}}
\label{eq:k1}
\end{equation}

\begin{equation}
log_{10}K_{\ref{eq:cr_k2}} = \frac{-14787}{T_{p}} + 4.5472 = \frac{X^{gas}_{CO} f^{1/2}_{O_2}}{X^{gas}_{CO_2}}
\label{eq:k2}
\end{equation}
where $T_p$ is the mantle potential temperature, and $X^{gas}_i$ is the mole fraction of species $i$ in the gas phase, and $f_{O_2}$ is the oxygen fugacity in units of pressure. At a given outgassing temperature, the values for $K$ and $f_{O_2}$ are constants, so these expressions can be solved for the species abundances given a mass-balance constraint. 

The specific outgassing rates for each gas are then given by normalizing the total carbon or hydrogen outgassing by the relative abundance of each gas species. The following equations are outgassing fluxes for C-bearing species and H-bearing species. To obtain the appropriate equations, you can substitute CO or CO$_2$ for $C_{species}$ in Eq. \ref{eq:rmorco}, and H$_2$ and H$_2$O for $H_{species}$ respectively in Eq. \ref{eq:rmorh}.

\begin{equation}
r_{Cspecies} = r_{outgas}^{C} \frac{X_{C_{species}}}{X_{CO_{2}}+X_{CO}}
\label{eq:rmorco}
\end{equation}

\begin{equation}
r_{Hspecies} = r_{outgas}^{H} \frac{X_{H_{species}}}{X_{H_{2}O}+X_{H_{2}}}
\label{eq:rmorh}
\end{equation}

All that is left is to specify the oxygen fugacity values. In this study, we adopt fixed mantle oxygen fugacities for the duration of any given simulation. For this study, we reference the Iron-W\"{u}stite (IW) buffer to define the range of oxygen fugacities we plan to explore. The oxygen fugacity of the IW reaction is calculated from the equation \cite{fegley2012practical}:

\begin{equation}
log_{10}f_{O_{2}} = 5.2614 - \frac{27,341}{T} + 0.4291 log_{10}T + \frac{0.0556 (P-1)}{T}
\label{eq:oxygfug}
\end{equation}
where $fO_2$ is in bars, $T$ is temperature in Kelvin, and $P$ is pressure in bars. We use IW as a reference point, and explore the range of oxygen fugacities from IW-2 (2 log units below the IW buffer) to IW+4 (4 log units above the IW buffer). This range explores reduced to oxidized conditions for the mantle oxidation state. Earth's upper mantle is roughly IW+4 (close to the Fayalite-Magnetite-Quartz or FMQ buffer at 1200 K). The interiors of both Mars and the Moon had initial mantle oxygen fugacities  close to IW-2 \cite{wadhwa2008}, although Mars' upper mantle likely oxidized over time and is now around IW+2 (FMQ-2) \cite{nicklas2021}. This range therefore spans most of the observed range within the Solar System rocky planets. 

\subsection{Initial Parameters} \label{sec:parameters}

Before we are able to apply this model to exoplanets, we validated it for Earth to ensure that our model was appropriately calibrated for the carbon and deep water cycles. To test the validity of the model, we matched three essential parameters to present day values for Earth: the heat flux (74.5 mW m$^{-2}$), upper mantle temperature (1600 K) and the amount of water on the surface (1.39 $\times$ 10$^{21}$ kg). We adopted the total mantle convective heat flux for the modern Earth  from \citeA{jaupart2015treatise}, averaged over the surface of the Earth. Additionally, we also tested different viscosity parameterizations to be able to obtain modern Earth. After testing the viscosity profiles from \citeA{sandu_2011} and \citeA{schaefer2015persistence}, we adopted the latter for our model because it better matched modern Earth values. In order to best match the modern Earth values, we varied the initial mantle temperature, the fraction of water in the mantle at the beginning of the simulation, and the total mass of water. Our best fit model for the Earth had 3 ocean masses worth of water, which is within the range of uncertainties on the Earth's total water content \cite{peslier2017water,hirschmann2018comparative,mccubbin2019origin,ohtani2021hydration}. Our best fit model also had an initial mantle temperature of 2300 K, with 55$\%$ of the initial water being stored in the mantle at the beginning of the simulation. We adopt these parameter values also for the TRAPPIST-1 planets.

As mentioned in Section \ref{sec:fO2}, we also test the effect of mantle oxygen fugacity on the composition and rate of outgassing. This can greatly influence the outgassing abundances of a planet. For each planet, we test oxygen fugacities at the iron-wustite (IW) buffer, IW - 2 and IW + 4. This range is equivalent to the range of oxygen fugacities that Earth's mantle has experienced since core formation through the present day \cite{rubie2015} and enables us to explore a range of possible outgassed atmospheres. Below, we discuss the parameter values we adopt for the TRAPPIST-1 planets and for the bulk planetary volatile abundances. 

\subsubsection{TRAPPIST-1 planets} \label{sec:trappist1}
We adopt nominal planetary mass and radius values for the TRAPPIST-1 planets from the analysis of \citeA{agol_2021}, as well as a nominal core mass fraction (CMF) based on their best fitting models. We calculate the equilibrium temperatures for these planets based on the received stellar insolation, assuming a planetary albedo of 0.3. For the mantle thermal evolution model, we also need to know the depth of the convecting layer (mantle), for which we need the radius of the core (R$_{core}$). We use the two-layer internal structure model of \citeA{Schaefer2017core} to calculate R$_{core}$ for each planet given R$_{p}$, M$_{p}$, and CMF, assuming a minimal radius fraction for the water layer. This model integrates the equations for mass and pressure from the center of the planet to the surface. We adopt a core of pure solid Fe and a mantle composed of magnesium silicates with appropriate phase changes. Note that the outcomes of the thermal evolution simulations are not particularly sensitive to R$_{core}$.
 
 \begin{table}[h]
 \caption{TRAPPIST-1 Planet Input Parameters}
 \centering
 \begin{tabular}{l c c c c c c c}%p{0.05\linewidth}}
 \hline
  TRAPPIST-1  & Radius & Mass & g & $T_{eq}$ & CMF & WMF (wt\%) & $R_{core}$ \\
  Planets & ($R_{\oplus}$) & ($M_{\oplus}$)  & ($g_{\oplus}$) & (K) & (wt\%) & (min, max) & ($R_{\oplus}$)\\
 \hline
  d & 0.788 & 0.388 & 0.624 & 262 & 19.7 & $10^{-3}$  & 0.38 \\
  &   &   &   &   &  & ($10^{-4}$, 2) &   \\
  \hline
  e & 0.92 & 0.696 & 0.817 & 228 & 24.6 & 0.3 & 0.48   \\
   &  &  &  &  &  & ($10^{-4}$, 2.1)  &    \\
   \hline
  f & 1.045 & 1.039 & 0.951 & 199 & 20.1 & 1.9  & 0.49   \\
   &  &  &  &  &  & (0.6, 3.4) &    \\
 \hline
 \multicolumn{8}{p{0.8\linewidth}}{\textbf{Note.} The radius, mass, gravity, and core mass fraction (CMF) were obtained from \citeA{agol_2021}. Equilibrium temperature and R$_{core}$ were calculated for the purposes of this paper.}
 \label{tab:planet_params}
 \end{tabular}
 \end{table}
 
\subsubsection{Volatile Abundances} \label{sec:volatiles}
In addition to the planet parameters, we test the influence of water mass fraction (WMF) on the outgassing abundances and therefore the atmospheric composition of these planets. The range of water mass fractions used for each planet are outlined in Table \ref{tab:planet_params}. For TRAPPIST-1e and f, we adopt a nominal water mass fraction for the 25\% CMF models of \citeA{agol_2021} and use the uncertainties on those WMFs to set an upper and lower value to explore. The lower uncertainty on WMF for TRAPPIST-1e predicts a value of zero for WMF, but to allow the model to run, we adopt a minimal value of 10$^{-4}$ wt\%. For TRAPPIST-1d, only an upper limit of 10$^{-3}$wt\% on WMF was determined for the nominal CMF by \citeA{agol_2021}, so to explore the effect of water abundance on our outcomes, we use a similar range to TRAPPIST-1e. To get the planets' carbon abundance, we assume that the planets have a chondritic H$_2$O/CO$_2$ mass ratio of 2.779 (calculated from elemental abundances in the Orgueil meteorite from \citeA{lodders2009abundances}), and calculate the CO$_{2}$ abundance for a given nominal water abundance. We adopt an initial distribution in which 55\% of the water is in the mantle, with the remainder at the surface, whereas we assume that 5\% of the carbon is initially in the mantle, 85\% in crustal reservoirs and the remaining 10\% in the atmosphere. The model is not strongly sensitive to these ratios, but they are roughly consistent with volatile distributions shortly after a magma ocean stage \cite{elkinstanton2008,zahnle2010}. 

\section{Results} \label{sec:results}

In the following subsections, we first describe the results of the outgassing model for the atmospheric compositions, surface temperatures, and surface ocean masses of the nominal models for TRAPPIST-1d, e and f. We then discuss variations in the results for each of the planets as a result of oxygen fugacity, followed by changes in the initial volatile abundances. We end with a discussion of the outgassing rates as a function of time. 

\subsection{Nominal Case}

\begin{figure}[t]
    \centering
%    \hspace*{-8mm}
    \includegraphics[width=\textwidth]{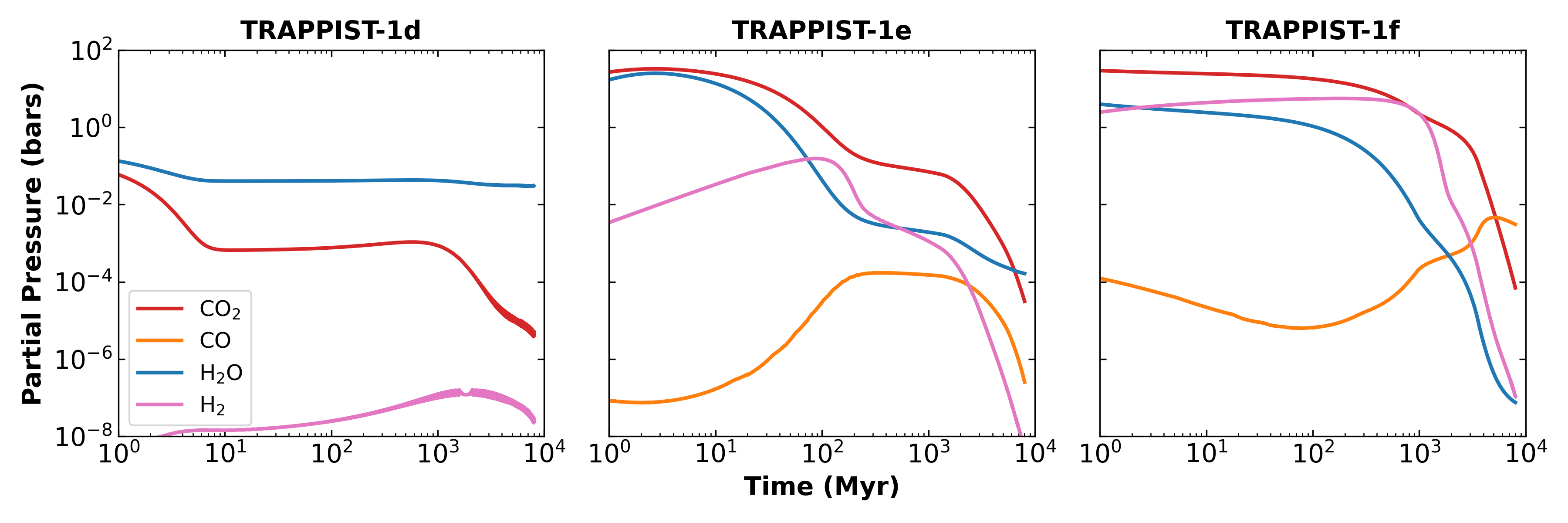}
    \caption{Partial pressure evolution of the atmospheric gases for the nominal cases for TRAPPIST-1 d (left), e (middle), and f (right). The oxygen fugacity is set to IW+4 and water mass fraction is set to a nominal value following \citeA{agol_2021}. CO$_{2}$ is given by the red line, CO by the orange line, H$_{2}$ by the pink line and H$_{2}$O by the blue line. We see that for the nominal cases, planet d is mostly dominated by H$_2$O and CO$_2$ over its evolution. Planet e and f have higher gas abundances for all gas species over their evolution with the difference that TRAPPIST-1e is dominated by CO$_2$ and H$_2$O and TRAPPIST-1f is dominated by CO$_2$, H$_2$ followed by H$_2$O. Planet f is the only one that has a CO dominated atmosphere at present-day. }
    \label{fig:spectracompare}
\end{figure}

\begin{figure}[h]
     \centering
%     \hspace*{-8mm}
     \includegraphics[width=100 mm]{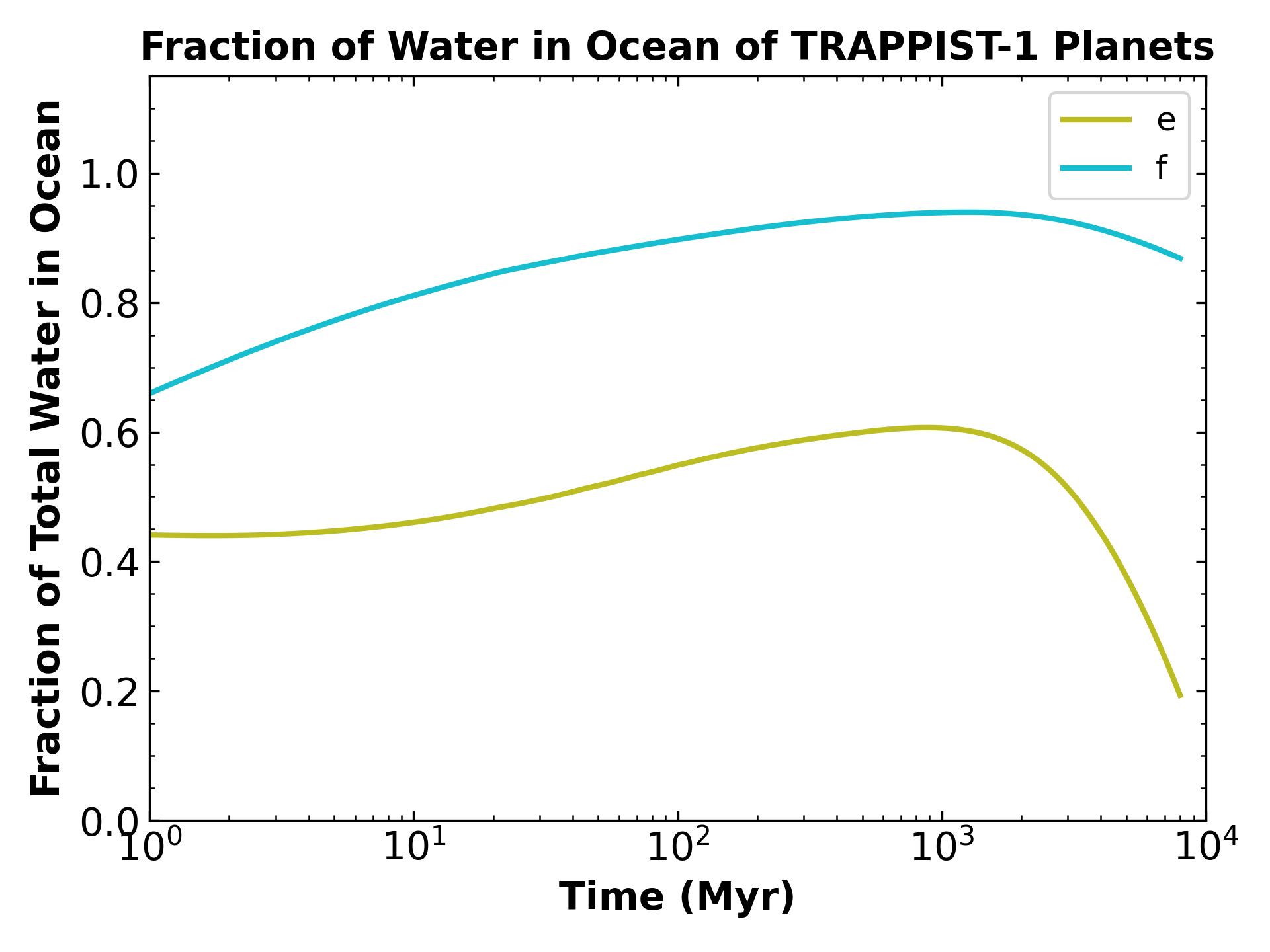}
     \caption{Evolution of the ocean mass for the nominal cases for TRAPPIST-1 e and f. Planet d is not shown because it does not have sufficient surface water to produce an ocean over its evolution. We normalize the ocean mass to its initial water abundance, which differs by planet. We observe that with increasing orbital distance from the star for the nominal cases, the outer planets are able to retain their water in liquid form on their surface for most of its evolution with planet e beginning to lose water to the interior around 1 billion years and planet f retaining the highest amount of liquid water at present day. Planet d has most of its water in the plate, with the remaining water surface mass in the atmosphere.}
     \label{fig:ocean_mass}
\end{figure}

\begin{figure}[h]
     \centering
%     \hspace*{-8mm}
     \includegraphics[width=100 mm]{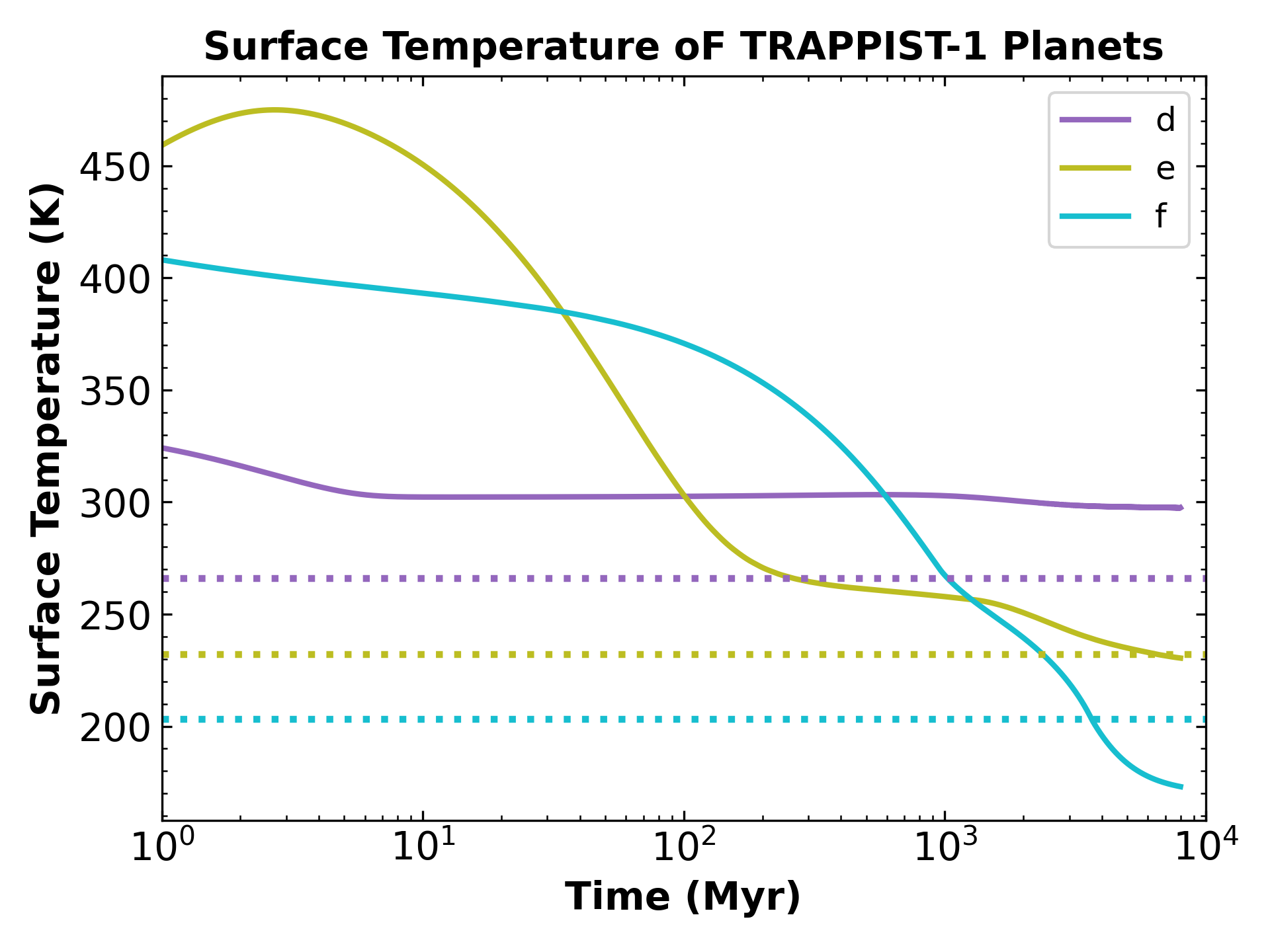}
     \caption{Surface temperature evolution of TRAPPIST-1d, e and f for the nominal cases over time. Planet d is given by the purple line, planet e by the olive-green line and planet f by the blue line. The dashed lines are their respective equilibrium temperatures. Planet e is the hottest planet of the three in the first 100 million years due to its high CO$_2$ abundance in its atmosphere and its equilibrium temperature, followed by planet f which also has a similarly high CO$_2$ abundance but a lower  equilibrium temperature. Planet d starts at the lowest temperature of all the planets, despite a higher equilibrium temperature, due to the low atmospheric pressures. However, at  present day, planet d is the hottest with its surface temperature exceeding the equilibrium temperature, followed by planet e reaching its equilibrium temperature and planet f being the coldest of all the planets with a surface temperature below its equilibrium temperature and significantly below the freezing point of water (273 K).}
     \label{fig:surf_temp}
\end{figure}

For all planets, we consider the nominal case to have an oxygen fugacity 4 log units above the iron-wustite buffer (IW+4), which is roughly equivalent to the oxygen fugacity of the Earth's present day upper mantle \cite{canil1997vanadium,delano2001redox,li2004constancy}. For this oxygen fugacity, we expect outgassing to produce volcanic gases similar to those seen on Earth, with water vapor and CO$_2$ dominating. We therefore expect these two gases to dominate the atmosphere composition for the nominal cases. The nominal WMF is given by the values in \citeA{agol_2021}, listed in Table \ref{tab:planet_params}. The nominal case for each of the TRAPPIST-1 planets is shown in Figures \ref{fig:spectracompare}, \ref{fig:ocean_mass}, and \ref{fig:surf_temp}, which show the time evolution of the atmospheric composition, the surface ocean mass, and the surface temperature, respectively.

\subsubsection{TRAPPIST-1d}

The nominal case for planet d (Fig. \ref{fig:spectracompare} \textit{left}) shows the atmospheric partial pressures for three of the four gas species as a function of time until the present age of the system ($\sim$7.4 Gyr, \citeNP{agol_2021}). Carbon monoxide has very low partial pressures throughout its evolution with abundances of less than one part per billion, which is the reason it is not shown in Fig. \ref{fig:spectracompare} (\textit{left} panel). Hydrogen is also maintained at low partial pressures compared to water vapor (H$_{2}$O) and carbon dioxide (CO$_{2}$), which are the dominant species in its atmosphere. 

Hydrogen builds up in the atmosphere very slowly, peaking in abundance at 100 parts per million around 2 billion years and slowly declining again. The amount of hydrogen and carbon monoxide is in general lower at the beginning of the simulation because we set their initial abundances to zero, while we assume an initial partitioning between the mantle and atmosphere for CO$_{2}$ and water to approximate an earlier magma ocean episode. For carbon dioxide, the partial pressure declines until around 10 million years, where it becomes stagnant until around 1 billion years. From there, there is another slow decrease in the CO$_{2}$ partial pressure until the end of the simulation. For water vapor, the partial pressure decreases slightly between 1 and 10 million years and it stays approximately constant at 0.04 bars through its later evolution stages. The maximum partial pressure is approximately 0.2 bars.

Broadly speaking, water vapor dominates the atmospheric composition of TRAPPIST-1d, which is different than the two other planets. This depends on the way that we partition the water in the surface reservoir between atmosphere, crust, and plate.  We assume that the atmospheric abundance of water remains at the saturated vapor pressure and that the remaining water is partitioned between the crust and ocean with water first going to hydrate the crust and the remaining water going to the ocean. The nominal case for d does not have enough water leftover to form an ocean, which is the reason planet d is not shown in Fig. \ref{fig:ocean_mass}. This procedure may underestimate the amount of liquid water present at the surface. This is also related to the surface temperature as seen in Fig. \ref{fig:surf_temp}, which is higher than its equilibrium temperature and keeps the saturated water vapor pressure high. Note that because liquid water is not present on the surface, we are slightly over-estimating water vapor pressure by adopting the saturated vapor pressure, but the effect on the return flux of water to the mantle is likely minimal. 

In Fig. \ref{fig:surf_temp}, we see that the surface temperature of planet d starts 1 million years is 335K and then decreases slightly to 302 K where it stays constant over time. At the current estimated age of the system, it remains well above its equilibrium temperature at 297K. This surface temperature is within habitable temperatures experienced in Earth's past, but since it retains most of its water in the plate and the rest in the atmosphere, it does not signal a habitable environment over its evolution. Our calculation adopts a perfect hydration efficiency for the crust, which would likely require the presence of liquid water to occur, so we may be overestimating the rate at which water partitions into the crust and therefore underestimating liquid water on the surface. Our model therefore cannot conclusively rule out a habitable environment on TRAPPIST-1d.

\subsubsection{TRAPPIST-1 e}

For TRAPPIST-1 e, CO has the lowest partial pressure, whereas CO$_{2}$ dominates the composition of the atmosphere throughout most of its lifetime. We note that the total atmospheric pressure (the sum total of the partial pressures of all of the gases) declines gradually over time. Compared with planet d, we see that CO$_{2}$ dominates instead of water vapor in its atmosphere. Planet e has initial partial pressures of 33 and 25 bars, respectively, for both CO$_{2}$ and H$_{2}$O in the first 100 million years, which decrease slowly over time until they reach atmospheric levels of roughly 3 $\times$ 10$^{-5}$ and 0.0001 bars, respectively, at present-day. While CO$_2$ is dominant throughout most of planet e's lifetime, water vapor is slightly more abundant than CO$_2$ from 2 billion years until present day.

Although both water vapor and carbon dioxide steadily decline over time, we note that both carbon monoxide and hydrogen start at partial pressures higher than those of planet d and increase for the first 100 million years before declining. Similar to planet d, CO is the least abundant gas in the atmosphere. However, CO is more abundant in planet e's atmosphere than planet d's, and it has a much larger peak abundance, going up to 0.0001 bars at 100 million years before decreasing at 2 billion years. This is due to having more water vapor, which essentially means have more OH present for the CO oxidation reaction to take place. When water vapor declines, CO increases in the atmosphere, which we can see directly in Figure \ref{fig:spectracompare}. Hydrogen is also significantly more abundant in planet e's atmosphere than planet d's, with a maximum value of 0.15 bars around 90 million years before decreasing steadily until present-day, ending with an abundance below that of CO at about 2 billion years. All gases decrease in abundance as volcanic outgassing declines over time.

Planet e starts with a significant ocean as shown in \ref{fig:ocean_mass}, which it retains for most of its lifetime. The ocean mass peaks at 0.55 fraction of the total initial water in the ocean around 1 billion years for the nominal case and then declines steadily over its lifetime. This is due to an increase in water being subducted into the mantle and outpacing the outgassing of water. At the present day, our nominal model predicts that TRAPPIST-1e has a surface ocean mass that is 0.04\% of the present-day Earth mass.

The surface temperature of TRAPPIST-1e begins at around 460 K at 1 million years where it shortly increases to 459 K due to small increases in atmospheric water vapor and CO$_2$ before decreasing until around 200 million years. This period of declining surface temperature also correlates with the decrease of the partial pressures of the water vapor and CO$_2$ as shown in \ref{fig:spectracompare}, which is driven by subduction. At around 200 million years, the partial pressures of H$_2$O and CO$_2$ continue to decrease but at a much slower pace, and as a result, the surface temperature of declines more slowly until around 2 billion years. Lastly, we see a sharper decrease in the surface temperature between 2 and 7.4 billion years, correlated with a sharp decrease in the CO$_2$ partial pressure. Of the three planets, planet e starts with the highest surface temperature, due to its relatively high volatile budget (higher CO$_2$ abundances than planet d) and higher equilibrium temperature than planet f. Both parameters combined lead to a higher surface temperature for TRAPPIST-1e. The surface temperature steadily declines as plate tectonics draws greenhouse gases out of the atmosphere, and it finally stabilizes near to the planet's equilibrium temperature at the end of the simulation when partial pressures of the greenhouse gases are all relatively low. The surface temperature remains below the freezing point of water after about 1 billion years, indicating that global glaciation may occur. We note that this model predicts a present-day atmospheric pressure for TRAPPIST-1e that is comparable to Mars. 

\subsubsection{TRAPPIST-1 f}

CO$_{2}$ dominates during most of TRAPPIST-1 f's evolution, with the CO$_2$ partial pressure beginning to decline around 1 billion years. We also see a slow increase in hydrogen over time with a much sharper drop-off around 1-2 billion years. For CO, the partial pressures is initially much lower than the rest of the gas species, it decreases slightly over time before starting to increase at around 100 million years, reaching a peak abundance of 0.004 bars at around 6 billion years. The behavior corresponds to CO being oxidized with increased water, while as water gets depleted in the atmosphere, less of it is available to react with CO. 

Despite a much higher total volatile budget, planet f showcases a similar initial partial pressure for CO$_{2}$ as planet e and higher initial partial pressures of H$_2$ than the other two planets at 5 bars. H$_2$O starts at a slightly lower partial pressure than planet e at 4 bars, but still considerably higher than planet d. CO$_{2}$ and H$_{2}$O are dominant around 1 million years, but hydrogen quickly surpasses water vapor in abundance. We note that H$_{2}$O is slightly lower than planet e at the present-day, reaching the lowest levels of water vapor of all three planets. Hydrogen is higher than planet d and e,  and it peaks around 1 billion years then declines sharply. The planet ends with an atmosphere where CO and CO$_2$ are more abundant than H$_2$O and H$_2$ at the present-day, although most of its evolution is dominated by CO$_2$, H$_2$ and H$_2$O. As with planet e, the total pressure of the planet's atmosphere at the present day is roughly equivalent to that of Mars. 

Planet f has a higher fractional ocean mass during its evolution than the other two planets. The fraction of water at the surface approaches 100\% of the total water budget in the ocean around 1 Gyr, where we see a very slow decline until the present day, where it has around 90\% of the total. Even though its surface is covered by water, we end up with less water vapor in the atmosphere because the planet is further away from its star so the saturation vapor pressure of water is lower. However, planet f ends up with a slightly larger surface ocean than it starts with, in contrast with the other two planets. 

The surface temperature for TRAPPIST-1 f begins at a lower temperature than planet e at 407 K that very slowly declines over the first few tens of million years before decreasing more steeply around 100 to 200 million years. At its current estimated age, we find the surface temperature is lower than its equilibrium temperature. The surface temperature evolution over time starting around 100 Myr closely resembles the trends of the water vapor partial pressures on the planet. The surface temperature drops below the freezing point of water at about 1 billion years, indicating that the planet may experience a global snowball state. We acknowledge that because the surface temperature drops below freezing point of water, it could potentially change the weathering efficiency and water partitioning in the planet, but we leave this for future studies. 

\subsection{Effect of Oxygen Fugacity}

Planetary interiors can affect the composition of outgassed volatiles through the mantle oxidation state, which can be quantified by the oxygen fugacity (fO$_{2}$) of the melt. Therefore, to account for a variety of mantle oxidation states, we use a wide range of oxygen fugacities. By accounting for the oxygen fugacity in the carbon cycle model, we hope to better understand its role and contribution to the outgassing rates and the atmospheric composition of these planets.

\begin{figure}[t]
    \centering
%    \hspace*{-8mm}
    \includegraphics[width=\textwidth]{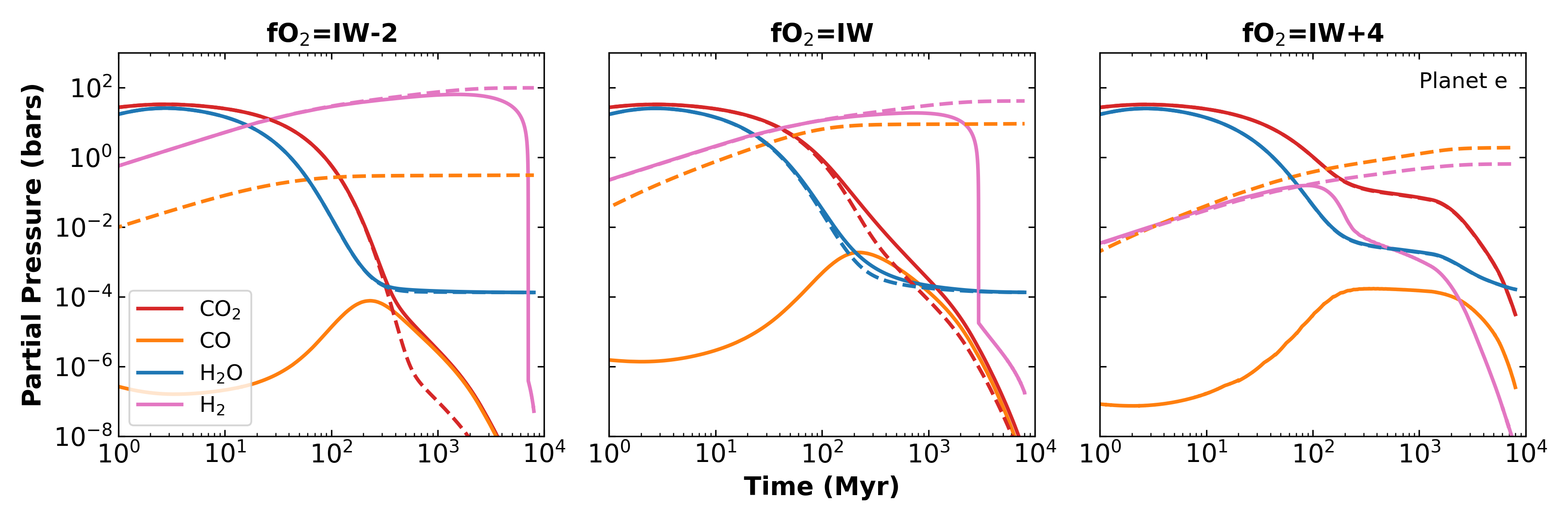}
    \caption{Comparison of the partial pressure evolution of the atmospheric gases for TRAPPIST-1e either with (solid lines) or without (dashed-lines) diffusion-limited escape and the CO-oxidation reaction. We assume the water mass fraction is the nominal value following \citeA{agol_2021}.  Left panel is for an oxygen fugacity of IW-2, middle panel is IW, and right panel is IW+4. CO$_{2}$ is given by the blue line, CO by the green line, H$_{2}$ by the pink line and H$_{2}$O by the orange line. We show both the effect of including CO oxidation reaction and diffusion limited escape to our model, which affects CO and hydrogen abundances significantly, as well as shows the the effect of oxygen fugacity on the partial pressures for TRAPPIST-1e.}
    \label{fig:planet_e_processes_off}
\end{figure}

We chose three oxygen fugacities for this study, which include the iron-wustite buffer itself (IW), in addition to 2 log units below (IW-2) and 4 log units above (IW+4) the buffer. These oxygen fugacities are used to probe both reduced and oxidized mantles for each planet. According to previous studies \cite{schaefer2017, kasting1993, guimond2023}, we expect to see more hydrogen (H$_2$) and carbon monoxide (CO) in planets with a lower oxygen fugacity (e.g. IW-2), which corresponds to reduced conditions. Similarly, we expect more water vapor (H$_2$O) and carbon dioxide (CO$_2$) in planets with a higher oxygen fugacity (e.g. IW+4), which corresponds to oxidized conditions. By exploring this range, we study the relationship between oxygen fugacity and the abundance of these different gas species in the atmosphere for the TRAPPIST-1 planets. In this section, we will discuss results for all three planets, with an emphasis on planet e (see Fig. \ref{fig:planet_e_processes_off}). Results for planets d and f are shown in Figures S1 and S2. To see the effect of the oxygen fugacity on the partial pressure of these atmospheric gases, we plot them over time until the present day for TRAPPIST-1 e in Fig. \ref{fig:planet_e_processes_off}. For all cases, we see all four gas species present in the atmosphere - CO$_2$, CO, H$_2$ and H$_2$O. 

For TRAPPIST-1e, in comparison to the nominal oxygen fugacity case (IW+4), we see that CO$_2$ and H$_2$O continue to dominate the atmospheric composition at the beginning of the simulation for both IW and IW-2, both reaching approximately 10 bars at 1 million years. For both of the reduced cases, CO$_2$ falls off more steadily and at earlier times than the IW+4 case, decreasing to negligible levels by 3 and 7 billion years, respectively. Water vapor has nearly identical behaviors for the two reduced cases, dropping to a relatively constant value of about 1 millibar around a few hundred million years, reaching slightly lower values than for the nominal case. Hydrogen is considerably more abundant in both reduced cases than the nominal case, reaching nearly 100 bars and 10 bars at IW-2 and IW, respectively. Both reduced cases show extremely sharp drop-offs in H$_2$ abundance, which occur at about 7 and 4 billion years, respectively. For both of these cases, H$_2$ is the most abundant gas for most of the simulation time, although H$_2$O becomes the most abundant gas at the very end of the simulations. For both of the reduced cases, CO remains the least abundant gas at all times, unlike the IW+4 case. Interestingly, the CO abundance is higher in the IW case, rather than the more reduced IW-2 case, reaching abundances of about 1 millibar in comparison to 0.01 millibars. This is due to the interplay of CO with water vapor through the CO oxidation reaction. In both reduced cases, CO peaks at about 100 million years and then drops steeply in a manner similar to CO$_2$. 

For TRAPPIST-1 d (Fig. S1), for all three oxygen fugacity cases, water vapor and CO$_2$ dominate during its evolution. Hydrogen begins at low partial pressures for all three cases and increases over 2 billion years, with much higher abundances for the two reduced cases, peaking slightly above and below $10^{-5}$, respectively, for the IW-2 and IW cases, compared to only $10^{-7}$ bars for IW+4. Lastly, the partial pressure of CO is so low during its evolution that we only start seeing values greater than a part per billion between 1 and 4 billion years for both the IW-2 and the IW case, and negligible values for the IW+4 case at all times.

 For TRAPPIST-1 f (Fig. S2), the major difference caused by reduced oxygen fugacity is that hydrogen becomes the most abundant gas, with abundances starting from 100-1000 bars and slowly increasing over its evolution. In contrast, for IW+4, hydrogen starts with a partial pressure between 1-10 bars, increasing over time to peak between 500 million years and 1 billion years before sharply declining. CO$_2$ and water vapor have similar abundances and evolution behavior in all three oxygen fugacity cases. Lastly, CO begins at low partial pressures compared to the other three gas species, with an early decline to a minimum abundance between 100-500 million years for all cases, before increasing to a peak abundance between 2-5 Gyrs. The highest initial abundance of CO is found in the IW case, but the IW+4 case has the highest final abundance of CO, with CO as the most abundant gas in the atmosphere at the end of the simulation. For IW-2 and IW, CO is the second most abundant gas after H$_2$ at the end of the simulations.

As described in Section \ref{sec:methods}, we added a simplified calculation for diffusion-limited escape of H$_2$ and for the CO-oxidation reaction in the atmosphere to our outgassing model. While we wanted to primarily see the effect of oxygen fugacity and initial volatile content on the outgassing rates for different gas species, we also wanted to test the contributions of these two processes to the atmosphere evolution. The contributions and differences in the partial pressures for the different gases can be seen in Fig. \ref{fig:planet_e_processes_off}.

For the simulations for planet e with these additional processes turned off, we see that the partial pressure of water in the atmosphere does not change. CO$_2$ also follows very similar trends for both cases with and without these additional processes. CO and H$_2$ are the gases that change significantly between the two cases, and both increase over time when these two additional processes are turned off in our model. We find that they dominate that atmospheric composition at the end of all of the simulations, including the oxidized IW+4 case, because all of the loss processes for these species have been turned off, whereas weathering and subduction fluxes continue to draw down H$_2$O and CO$_2$. Models that do not account for sinks for the reduced gases will over-predict their abundances in rocky planet atmospheres. 

For the figures for planet d and f, please refer to the supplementary figures S10 and S11. For planet d, both H$_2$ and CO increase and deviate significantly from the partial pressures we obtained with the CO-oxidation reaction and diffusion-limited escape of H$_2$ included. When these two processes are not in the model, the partial pressures of H$_2$ and CO surpass the partial pressure of CO$_2$ at modern day for all oxygen fugacity values used. For the lower oxygen fugacity case (IW-2) with these processes not included, H$_2$ matches the partial pressure of water vapor (H$_2$O) at present day at 5 $\times 10^{-1}$ bar. For the nominal case for oxygen fugacity (IW+4), CO becomes the second most dominant gas followed by H$_2$ when sinks are turned off, which contrasts with the lower oxygen fugacity cases. We observe the same behaviors for planet f when sinks are turned off - CO becomes more abundant than H$_2$ at modern day for the nominal oxygen fugacity case. For the two lower oxygen fugacity cases, neither CO$_2$ nor H$_2$O are affected, but CO becomes the second most abundant gas throughout most of the evolution. H$_2$, which is the most abundant gas for planet f at these oxygen fugacities for the nominal model, remains the most abundant gas when escape is turned off. 

\subsection{Effect of Water Abundance}

\begin{figure}[t]
    \centering
%    \hspace*{-8mm}
    \includegraphics[width=\textwidth]{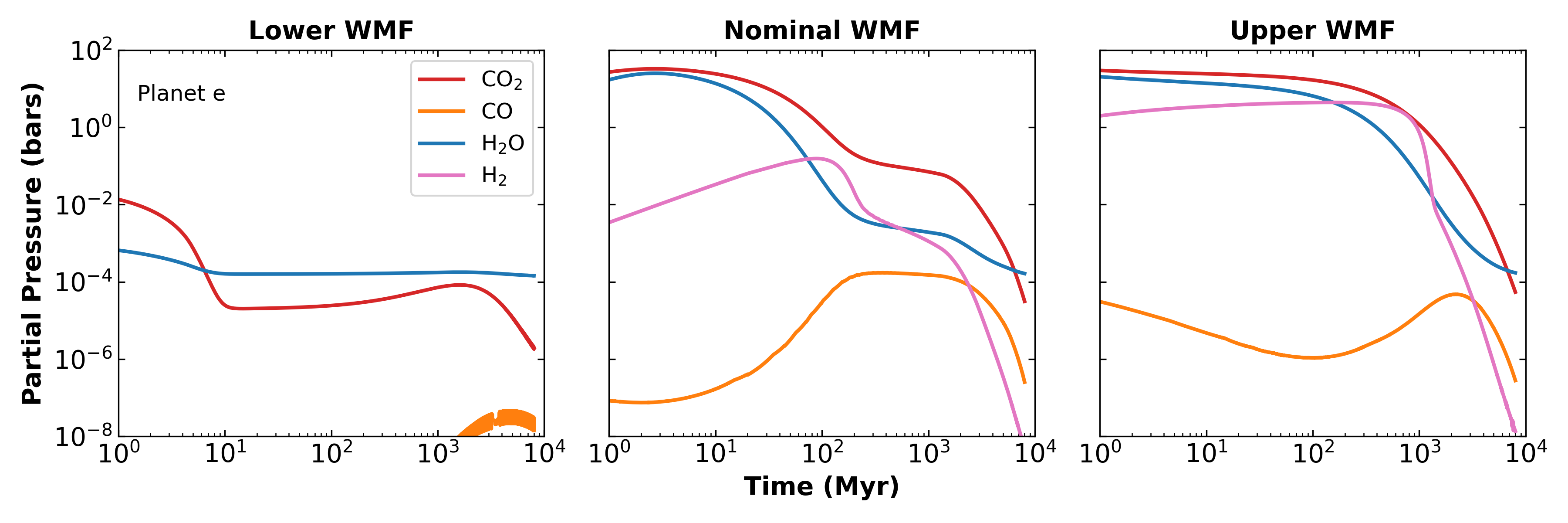}
    \caption{Comparison of the partial pressure evolution of the atmospheric gases for TRAPPIST-1e with different initial water mass fractions. We assume the oxygen fugacity has the nominal value of IW+4.  Left panel is low WMF, middle panel is the nominal value (same as in Fig. \ref{fig:spectracompare}), and right panel is high WMF. CO$_{2}$ is given by the red line, CO by the orange line, H$_{2}$ by the pink line and H$_{2}$O by the blue line. We see a significant effect of the WMF on the overall abundances of these gas species with lower water content leading to less outgassing, while higher water content leads to a higher abundance of most gases. The thickness in the CO line for the lower WMF is due to numerical noise, so we focus on the trend and the overall abundance for interpretation purposes.}
    \label{fig:planet_e_diff_WMF}
\end{figure}

The initial volatile abundances of a planet can influence its evolution and atmospheric composition. Therefore, we study the effect of different initial water abundances for each planet using the range given in \citeA{agol_2021}. This parameter is specifically indicated here as water mass fraction (WMF), which sets the initial amount of water in the planet at the beginning of our simulations (see Table \ref{tab:planet_params}). We assume that carbon co-varies with water at the CI chondritic relative abundances. In Fig. \ref{fig:planet_e_diff_WMF}, we show the effect of initial water abundance on the atmospheric composition of TRAPPIST-1 e. For the other two planets, please refer to the supplementary material. 

As with oxygen fugacity, we compare the effect of low and high WMF on the partial pressures of all of the gases in the atmosphere to the nominal case for TRAPPIST-1 e in Fig. \ref{fig:planet_e_diff_WMF}. At the lower WMF case, we see that the abundances of all of the gases in the atmosphere are significantly lower than our nominal case. While water vapor initially has the second highest partial pressure, it dominates in the atmosphere for most of its evolution at a near constant value of 0.001 bars, which is lower than its minimum pressure in the nominal case. CO$_2$ starts at 0.01 bars and decreases over 10 Myr to under $10^{-4}$ bars. It remains near this level until 3 Gyr before decreasing again. CO does not reach significant partial pressures in the atmosphere until 700 Myr with a peak abundance near the end of the simulation of about $10^{-8}$ bars. Hydrogen is not present in the lower WMF case for planet e.

In contrast, for the higher WMF case, all four gas species present. However, CO$_2$, H$_2$O and H$_2$ are particularly dominant throughout most of its evolution, with CO being the least abundant gas as expected. Both CO$_2$ and water vapor begin with partial pressures over 10 bars. CO$_2$ remains constant until 300 Myr where it begins to steadily decrease. Water vapor follows a similar trend but decreases slightly earlier at around 100 Myr. Water vapor partial pressures plateau around $10^{-3}$ bars at the present age, becoming the most abundant gas at the end of the simulation. Hydrogen follows closely behind, starting out over 1 bar and slightly increasing over time before sharply decreasing at 1 Gyr. At its current age, it is the least abundant gas followed by CO. Lastly, CO decreases during half of its evolution before increasing around 200 Myr and peaking at approximately $10^{-4}$ bar. 

Planet d with the lower WMF case exhibits only two gases dominating the atmosphere - H$_2$O and CO$_2$. Water vapor remains near 0.1 bars throughout its entire evolution, which is slightly lower than in the nominal case. CO$_2$ shows similar time-evolution behavior, but has abundances of one to two orders of magnitude lower than the nominal case, starting at approximately 0.01 bars, and decreasing to around 10$^{-5}$ bars until 1 billion years, followed by a slight increase and peaks near 10$^{-4}$ at 3 billion years. Interestingly, the final CO$_2$ abundance is very similar to the nominal WMF case. CO and H$_2$ are not present at significant abundances for the lower WMF case. For the higher WMF case,  both H$_2$O and CO$_2$ are nearly 3 orders of magnitude more abundant than in the nominal case, reaching up to 100 and 20 bars, respectively. Water vapor decreases to approximately  0.1 bars by modern age, consistent with the nominal case, whereas CO$_2$ decreases to a partial pressure of 10$^{-4}$ bars by the current age, about one order of magnitude larger than the nominal case. H$_2$ is initially more than eight orders of magnitude more abundant than in the nominal case, starting at 1 bar and slowly increasing over 1 Gyrs before sharply decreasing to 10$^{-5}$ bars, about 3 orders of magnitude higher than the nominal case. CO has the lowest partial pressure over time of the four gas species decreasing over the first 100 million years and increasing after that to a peak of 10$^{-5}$ bars. After 1 billion years it decreases to very low levels by the current age. However, this is still considerably more than the nominal case, for which the CO pressure never exceeds 10$^{-8}$ bars. In none of these cases is planet d able to retain liquid water on its surface. 

For TRAPPIST-1 f, the initial abundances and early evolution of both H$_2$O and  CO$_2$ are very similar for the three different cases. The abundances of these two gases at the present-day, however, seems to differ by WMF, with the lower WMF leading to a lower partial pressure at modern day compared to the higher WMF case having a higher partial pressure. For H$_2$, the higher the WMF, the more hydrogen there is both at the beginning of the simulation and at the current age. Carbon monoxide behaves slighlty differently, with higher initial abundances with increasing WMF, but decreasing present-day abundances. For both the nominal and lower WMF cases, CO becomes the most abundant gas species by the end of the simulation, but it remains the least abundant species for the high WMF case. This is likely due to the change in late-stage water vapor partial pressures, since water vapor plays a key role in the CO-oxidation reaction.  

\subsection{Outgassing Rates}

\begin{figure}[h]
    \centering
%    \hspace*{-8mm}
    \includegraphics[width=\textwidth]{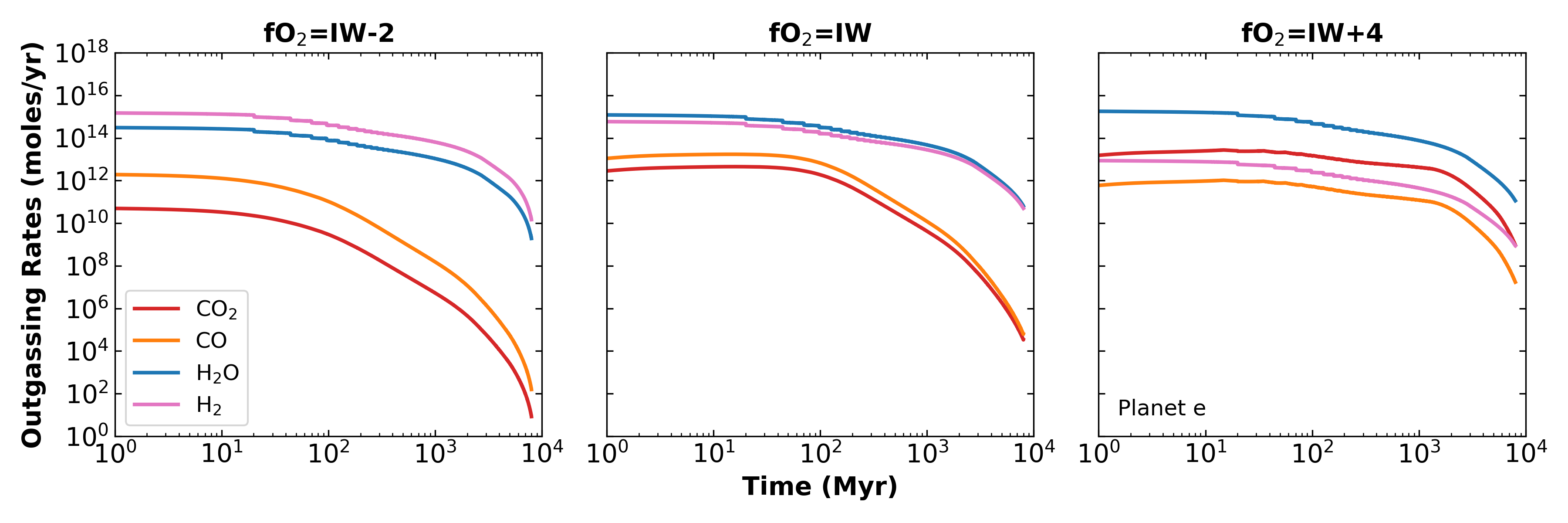}
    \caption{Comparison of the effect of oxygen fugacity on the outgassing rates of the different gas species for TRAPPIST-1e.  Left panel is for an oxygen fugacity of IW-2, middle panel is IW, and right panel is IW+4. CO$_{2}$ is given by the blue line, CO by the green line, H$_{2}$ by the pink line and H$_{2}$O by the orange line.}
    \label{fig:planet_e_outgassing_rates}
\end{figure}

One feature we highlight that we report here from our model compared to other geological evolution models in the literature \cite{krissansen-totton2022, krissansen-totton2024erosion, thomas2025} is the outgassing rates for all four gas species in this study (CO$_2$, H$_2$O, CO and H$_2$). These rates explicitly come from the various geological processes included in this model from both the carbon cycle and the deep water cycle. In Fig. \ref{fig:planet_e_outgassing_rates}, the outgassing rates are plotted over time for TRAPPIST-1 e while varying the oxygen fugacity. This allows us to compare how the outgassing rates (or source fluxes) compare with the actual atmospheric abundances shown in Fig. \ref{fig:planet_e_processes_off}, which are governed additionally by other sink and source processes. In addition, it is important to report outgassing rates because many photochemical models, which compute the influence of shorter timescale processes than our evolution model can account for, must adopt source and sink fluxes for the different gases at the surface, including estimates of volcanic outgassing. These results can therefore serve as constraints for photochemical models at different stages of the planets evolution. For the respective plots for planets d and f, please see the supplementary material.

In Fig. \ref{fig:planet_e_outgassing_rates}, we see that water vapor (H$_2$O) and hydrogen (H$_2$) is the most outgassed species over the planet's evolution for all oxygen fugacities, although the rate of outgassing is larger for higher oxygen fugcaities. Outgassing of H$_2$ is comparable to H$_2$O at the lowest oxygen fugacity, but declines at higher fO$_2$. The carbon species both have low outgassing rates at IW-2, but increase with increasing fO$_2$. CO dominates the carbon outgassing at both IW and IW-2, but CO$_2$ has higher outgassing rates for IW+4. The trend in carbon outgassing rates with oxygen fugacity is consistent with lower carbon solubility in the melts at low fO$_2$ and the formation of graphite. At IW+4, water vapor and CO$_2$ dominate the outgassing fluxes, consistent with what is seen on Earth, which has an upper mantle at a similar fO$_2$. We note that all of the outgassing rates decrease over time, with peak outgassing rates occurring at the beginning of the simulations, when the planet's interior is the hottest. 

Comparing these outgassing rates to the partial pressures of planet e for the same oxygen fugacity (see Fig. \ref{fig:planet_e_processes_off}), we note that hydrogen and CO abundances generally increase over time in the atmosphere despite the decrease in outgassing rates over the same amount of time. Comparing IW and IW-2, the higher H$_2$ and lower CO pressures for IW-2 are consistent with the relative outgassing rates in these simulations. The CO outgassing rates are actually relatively similar in magnitude for the IW-2 and IW+4 cases, which is consistent with the similar atmospheric pressures found. In contrast, the H$_2$ outgassing rate is lowest at IW+4, which is consistent with that case having the lowest atmospheric partial pressure. Despite having similar outgassing rates at IW and IW+4, H$_2$ always has higher atmospheric partial pressures than CO due to the efficiency of the CO-oxidation reaction, which converts CO into CO$_2$. Both H$_2$O and CO$_2$ partial pressures decrease over time similar to their outgassing rates, but we find slighlty higher CO$_2$ abundances at IW+4, where the CO$_2$ outgassing rate is much higher than at IW or IW-2. In contrast, the final water vapor abundance is similar across all of the oxygen fugacities and water vapor is never the most abundant species in the atmosphere, despite having the highest outgassing rates in all cases. This is because condensation of water removes most of it from the atmosphere, so its atmospheric partial pressure is dictated by the saturation vapor pressure at the surface temperature.

\begin{figure}[t]
    \centering
%    \hspace*{-8mm}
    \includegraphics[width=\textwidth]{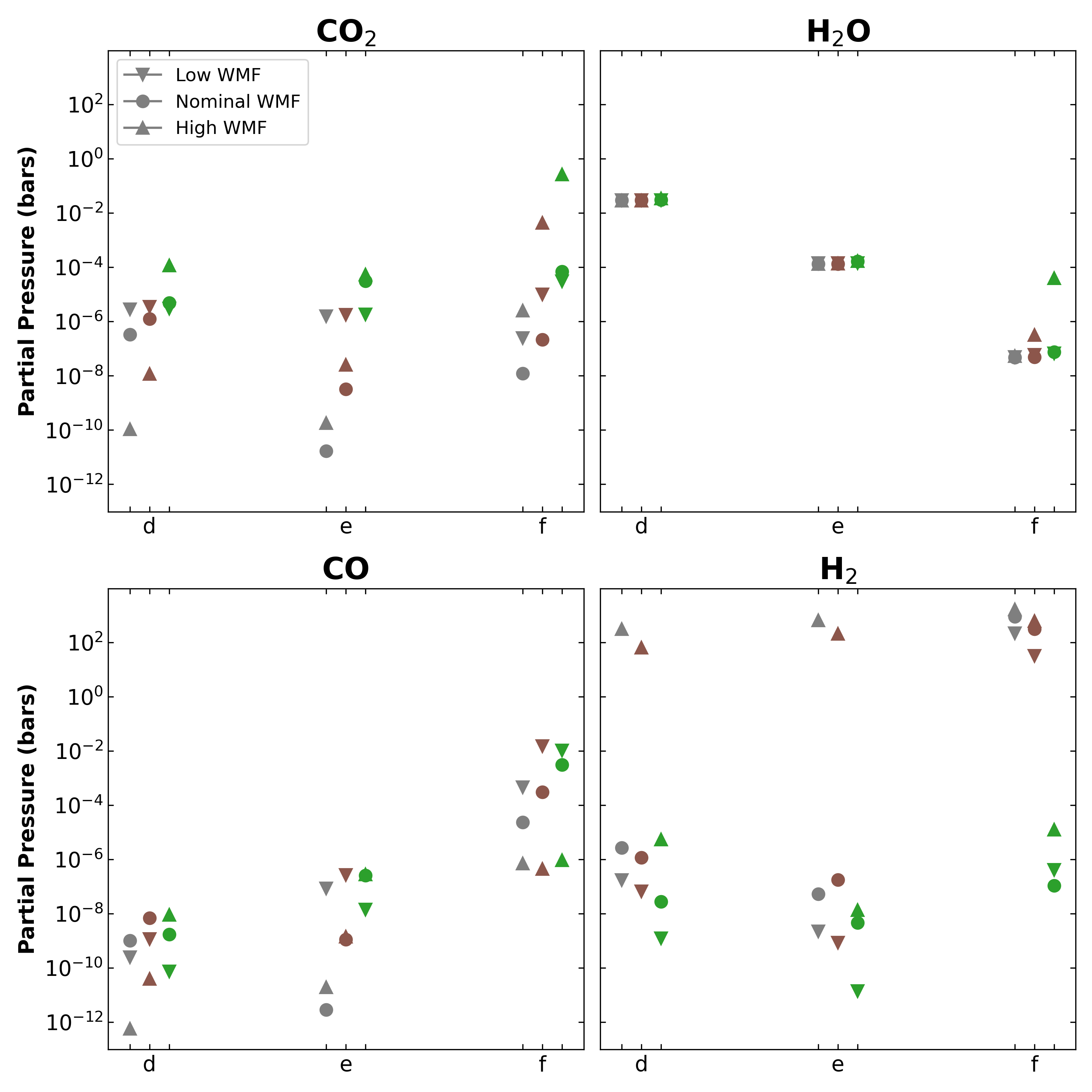}
    \caption{Summary figure showing the calculated partial pressures of the different gas species included in the model at present day for all three planets given the different oxygen fugacity and water mass fractions. Gray markers show IW-2, brown markers show IW and green markers show IW+4. The downward triangle is the lower WMF, the circle is the nominal WMF and the upwards triangle is the higher WMF cases for each planet.}
    \label{fig:summary-fig}
\end{figure}

\subsection{Final Atmosphere Compositions} \label{sec:finalabundances}

In Figure \ref{fig:summary-fig}, we show the present-day estimated partial pressures for the four gas species for the TRAPPIST-1 planets. For CO$_2$, we see a strong dependence on oxygen fugacity, with abundances varying by about 4 log units, which is stronger for the higher water mass fraction cases. For H$_2$O, we do not see any dependence on the oxygen fugacity or water mass fraction except for planet f for the higher oxygen fugacity case where we see a small difference. This is likely due to the fact that we calculate the saturated water vapor pressure, which depends most strongly on surface temperature. For CO, there is about 2 orders of magnitude variation in pressure due to oxygen fugacity, in comparison to 3-4 orders of magnitude variation due to the water mass fraction. Lastly, H$_2$ shows a very large dependence on water mass fraction, especially at low oxygen fugacities, of up to 8 orders of magnitude. The dependence on oxygen fugacity is moderate for the low and nominal WMF cases, whereas there is a very large dependence on oxygen fugacity for the high water mass fraction cases.  Using this, we can see that the effects of these parameters on the atmospheric abundances are not strictly linear, and that exploring a range of parameters for both oxygen fugacity and volatile abundances are necessary to see the range of atmospheric compositions that could exist.

\section{Discussion} \label{sec:discussion}

\subsection{Comparison with Observations}

No definitive detections of atmospheres on any of the TRAPPIST-1 planets have yet been made, but a number of observations have helped to rule out different classes and masses of atmospheres for some of the planets. \citeA{dewit2018atmospheric} osberved two mutual transits of TRAPPIST-1d, e, f, and g with HST/WFC3 and ruled out clear H$_2$-dominated atmospheres for all four planets, but was unable to make any constraints on heavier gases. Recent JWST observations of TRAPPIST-1e find that stellar contamination dominates the transmission spectrum taken with NIRSpec PRISM, but the studies are able to rule out H$_2$-dominated atmospheres and find that an N$_2$-rich atmosphere with trace CO$_2$ or CH$_4$ are allowed, but may not be fully consistent with all spectral features \cite{espinoza2025jwst,glidden2025jwst}. The low preferred CO$_2$ abundances in the best-fit spectra are consistent with the low atmospheric pressures that we find for all present-day models of TRAPPIST-1e. Likewise, observations of TRAPPIST-1d with NIRSpec/PRISM find significant stellar contamination and rule out clear H$_2$-atmospheres and thick atmospheres with high mean molecular weights \cite{piaulet2025strict}. Our models predict that water vapor dominates the atmosphere of TRAPPIST-1d, with pressures of 10$^{-2}$ bars, which is consistent with the derived constraints. They also find CO$_2$ pressures must be 10$^{-4}$ or less, which is consistent with our final pressures of $\sim$10$^{-6}$ bars, or 10$^{-4}$ for our high WMF case. We note that our high WMF case for TRAPPIST-1d has $\sim$33 oceans of total water budget, but remains consistent with the observations, indicating that tectonics alone may drive low surface pressures, rather than requiring significant atmospheric escape.  

\subsection{Model Assumptions and Limitations}

Our model used for this study makes various assumptions, which we will discuss in depth here. These models can serve as a way to understand how different interior-surface interactions could impact the atmosphere. It is important to note that there are many evolution pathways that a planet could take after formation, and this investigation serves as a starting point for understanding the geological evolution of rocky worlds around M-dwarfs stars. We do not claim that any of these planets have a specific oxygen fugacity or water mass fraction, but rather aim to explore parameter space and provide geological constraints and context to atmospheric observations. 

Our model is based on our knowledge of Earth's geological processes to date. One of the biggest assumptions in our model is that plate tectonics is active throughout our simulations. Although the onset of plate tectonics on Earth is thoroughly debated in the literature \cite{stern2005evidence,palin2020secular,cawood2018geological}, we assume that plate tectonics are active throughout the entire evolution of the planet. This study mainly serves as a first look at the effect of plate tectonics, geological cycling, and different interior parameters on the atmospheric composition of the planets. Adopting a stagnant-lid or episodic plate tectonic model could be considered in future studies, but is beyond the scope of the present study. However, we note that a stagnant lid model  would essentially mean that neither subduction nor volcanism along plate boundaries would be taking place. While outgassing through a stagnant lid and weathering could take place \cite{noack2017volcanism,dorn2018outgassing,foley2018carbon,tosi2017habitability}, this would mean there is no efficient cycling of volatiles from the mantle to the surface and the atmosphere.  Essentially the ingassing terms for all the mass-balance equations would change and essentially negligible for the model. The cycles would become stagnant and atmospheric processes would mostly dictate the behavior of the different gas species. In an episodic plate tectonic regime it would be similar to the stagnant-lid regime, but with the difference that plate tectonics would take over for a period of time which can move volatiles within the layers but might not be long or efficient enough to be able to retain similar partial pressures as the tectonic regime. 

One of the parameters we assume in our model is the initial water abundance by using the \citeA{agol_2021} values for the TRAPPIST-1 planets, which we discuss more below. We explore a wide range for the water mass fraction as described in \citeA{agol_2021}, but note that strong constraints do not exist for most other exoplanets. We also assume an initial carbon abundance in the planet by assuming a chondritic ratio between initial water and carbon abundances in the planet, similar to Earth. It is certainly possible for water and carbon abundances to vary from a chondritic ratio: neither Earth nor Mars have chondritic H/C ratios, for instance. However, varying water and carbon independently opens a wide parameter space, which we will explore in future work. 

From our initial carbon abundance, our model then assumes CO$_2$ begins with 5 percent of it in the mantle, 85 percent in the plate and 10 percent in the atmosphere. Note that while we do not simulate an earlier magma ocean phase, these fractions would be consistent with a magma ocean outgassing period followed by a period of rapid sequestration of CO$_2$ into the plate, as found by \citeA{sleep2001carbon}. Different percentages were tested before assuming these ratios, and this can be a place for further consideration in the future as we expand the number of exoplanets explored. We also assumed the initial hydrogen and carbon monoxide abundances to be zero at the beginning of the simulation. This explains why the partial pressures for both of these gas species are lower at the beginning of the simulations for all the TRAPPIST-1 planets in this study for all cases when oxygen fugacity and water mass fraction is varied. If H$_2$ and CO would begin at a higher percentage in the mantle, this would mean a quicker accumulation of these species in the various reservoirs included in this model rather than these species resulting from the various geochemical processes included in the model first before being present in the system. This assumption that can also be explored further in future studies.

Other initial parameters in our model include the initial percentage of water that is in the atmosphere and in the mantle, and the initial temperature of the mantle. All of these parameters were explored in depth by running a grid of simulations as described in Section \ref{sec:methods} to validate the model with Earth. The values that gave the best fit to Earth parameters were an initial mantle temperature of 2300K and 55 percent of water beginning in the mantle and 45 percent in the atmosphere. We therefore assumed these values for the TRAPPIST-1 planets, but note that these parameters can also be varied and explored in future studies. We also assumed an albedo of 0.3 in our study for the TRAPPIST-1 planets, but highlight the importance of exploring this parameter in the future \cite{rushby2020}. The TRAPPIST-1 planets could have lower albedos, which could lower their equilibrium temperatures \cite{wolf2025}. 

As mentioned above, our model focuses on the geological evolution and geological cycling during solid state convection, and we do not account for a magma ocean phase in this study. In future work, we plan to couple this model with an existing magma ocean model \cite{schaefer2016predictions}. The magma ocean phase can help set the initial volatile partitioning between mantle and atmosphere, which we have arbitrarily set in this study. 

We also assumed a mantle oxygen fugacity buffer that remains the same throughout the entire planet's geological evolution. However, the oxygen fugacity of Earth's early mantle is currently debated and could have evolved over time. In future work, we plan to modify this model to additionally track the evolution of the mantle oxygen fugacity, due to production of oxidants at the surface, metal production in the deep mantle and loss to the core, and partial melting processes.  

This study also highlights the importance of not only accounting for geological processes for the resulting secondary atmospheres, but also accounting for atmospheric processes such as oxidation reactions and diffusion-limited escape of H$_2$. These two processes are a relatively simple and minimal way to incorporate interactions within the atmosphere in our model, but we stress that photochemistry models are needed to provide better constraints on the abundances of the gases in the atmosphere after these interactions. In  Fig \ref{fig:planet_e_processes_off}, we see the effect of including these two processes for TRAPPIST-1 e. Without accounting for these processes, our model would predict that planet e's atmosphere would be hydrogen and CO dominated at present age. But by adding a CO-oxidation reaction and adding the diffusion-limited escape of H$_2$ we see that most of the CO and H$_2$ that are outgassed and retained in the atmosphere decrease over time. CO is used up completely through the CO-oxidation reaction and while hydrogen is gained with the oxidation reaction, it is simultaneously lost to diffusion-limited escape. In contrast, water and CO$_2$ have a very small change when we modify whether the two atmospheric processes are included or not. 

\subsection{Volatile abundances of the TRAPPIST-1 planets}

While we have adopted the present day abundances derived from \citeA{agol_2021}, we note that other internal structure models have found different WMFs. \citeA{boldog2024} used an internal structure model to estimate the water content of a number of rocky exoplanets, including TRAPPIST-1d, e, and f. They consider a planetary structure with separate liquid ocean and high pressure ice polymorph layers. They find water mass fractions from 2 – 21\% would satisfy the internal structure model constraints for all of the TRAPPIST-1 planets, unlike \citeA{agol_2021}, who find that the inner planets likely have much lower water abundances. We note, however, that \citeA{boldog2024} do not state the corresponding iron core mass fractions for the different model solutions, or whether those values are within physically plausible ranges. Another recent reassessment of the internal structure of TRAPPIST-1f suggests that it may be even more water-rich than found by \citeA{agol_2021} \cite{rice2025uncertainties}. Their preferred WMF is $> 16 wt\% $for TRAPPIST-1f, which is well above the maximum value we consider here of 3.4 wt\%. Because our model does not independently model the water layer (e.g. ocean depth, ice-layer formation), it would not be suitable to use for a planet with 16 wt\% water, so our results for this planet must be taken with caution. 

A number of papers have also looked at how the total water budget of the TRAPPIST-1 planets has evolved to the present day values through processes such as atmospheric escape. Our models assume a static water abundance because although we allow H escape, we do not find considerable mass loss, and we assume that our evolution model begins after the end of the magma ocean phase, during which most of the hydrodynamic escape occurs. Using a steam atmosphere--magma ocean model, \citeA{barth2021magma} find substantial loss may have occurred for both TRAPPIST-1e and f, with TRAPPIST-1e becoming fully dessicated for low initial volatile inventories consistent with our nominal case, but losing less than 20\% of the initial volatile abundance if the planet begins with more than 50 Earth oceans, equivalent to 1.7 wt\% water. In contrast, TRAPPIST-1f retains more than 20\% of total water for initial budgets greater than about 10 oceans (~0.22 wt\%), which is lower than our lowest WMF case. \citeA{barth2021magma} do not report results for TRAPPIST-1d. More recently, \citeA{van2024airy} calculate present day upper atmosphere structures and thermal Jeans escape rates for the TRAPPIST-1 planets and find that all of the TRAPPIST-1 planets could lose 1 bar of atmosphere in about 1 Myr, with escape rates that would only be higher in the past when the stellar XUV was stronger. This indicates that significant mass loss is likely to have occurred from all of the TRAPPIST-1 planets. \citeA{gialluca2024implications} incorporated uncertainties on stellar parameters into a study of hydrodynamic escape from the TRAPPIST-1 planets. They find that TRAPPIST-1d would be totally dessicated if it started with less than 50 Earth oceans of water (3 wt\% of the planet), whereas maximum water losses for TRAPPIST-1e and f were 8 and 4.8 Earth oceans, respectively, equivalent to 270 ppm and 100 ppm of planetary inventory. This model assumes that all water is on the surface and available to escape from a pure steam atmosphere very early in the star’s history, and therefore maximizes total water loss through hydrodynamic escape. Based on their escape models and the current inferred water abundances, they derive initial water abundances of 90 and 200 Earth oceans for TRAPPIST-1e and f, respectively, equivalent to 3 wt\% and 4.5 wt\% of the planets. These values are slightly higher than the maximum WMF that we take for TRAPPIST-1e and f of 2.1 and 3.4 wt\%, respectively.  

\subsection{Comparison with other models}

Although our study uses and improves an existing atmosphere-interior model, similar models exist to simulate atmosphere-interior interactions. In this section, we will briefly share the differences between our model and other existing models that have been explicitly used to study the TRAPPIST-1 planets. We do not discuss other atmosphere-interior models that have not been used to study the TRAPPIST-1 planets, but highlight they could potentially be used in the future for similar studies.

Krissansen-Totton and colleagues have developed the Planetary Atmosphere, Crust and Mantle evolution (PACMAN) model over several papers with application to the TRAPPIST-1 planets \cite{krissansen-totton2021venus,krissansen-totton2021waterworlds,krissansen-totton2022}. In an early version of the model, \citeA{krissansen-totton2021waterworlds} uses a melt solubility model to show that overburden pressures at the base of a large global ocean may be sufficient to suppress outgassing from subaqueous volcanic sources. They find that pressures greater than $\sim$1 GPa are sufficient to prevent outgassing of most volatiles. They estimate vent pressures on the outer TRAPPIST-1 planets using the same volatile abundances from \citeA{agol_2021} that we use here and find that outgassing is unlikely to occur on TRAPPIST-1f and g. We do not include the effect of overburden pressure on outgassing rates in our models. For TRAPPIST-1f, we find that the ocean mass produced in our nominal case at IW+4 produces pressures of 0.9 - 1.8 GPa, which indicates that our model may overestimate outgassing for that set of planet parameters. However, for our low WMF case, we find that pressures at the base of the ocean stay well below this critical threshold and outgassing is likely to continue over the planet's lifetime. For our lowest oxygen fugacity case (IW-2), the nominal and low WMF cases also stay below this critical pressure threshold, so we find that outgassing could be possible for TRAPPIST-1f under low to moderate WMF or low oxygen fugacity conditions. 

The full PACMAN model \cite{krissansen-totton2022,krissansen-totton2023nondetection} simulates planetary evolution for terrestrial planets from the magma ocean phase to the solid-state geochemical cycling. Processes include hydrogen escape, outgassing of gases, oxidation of new crust, weathering, among others. It can also simulate plate tectonics or stagnant lid in the solid-state phase. A major difference with our model is that PACMAN includes both the magma ocean phase and the solid-state geochemical cycling, while ours only models the latter. While a magma ocean phase can be added to the model used for this study, it is outside of the scope of this study. When using PACMAN, a plate tectonics regime was assumed which is the same as this study. In \citeA{krissansen-totton2022}, PACMAN is able to calculate outgassing fluxes for CO$_2$, H$_2$O, CO, H$_2$ and CH$_4$, but only specifically tracks CO$_2$, H$_2$, and O$_2$ in the mantle and surface of the TRAPPIST-1 planets due to the assumption that reductants (CO, H$_2$, CH$_4$) instantaneously deplete oxygen in the atmosphere. In contrast, our model both calculates and tracks CO$_2$, CO, H$_2$O, and H$_2$ due to including the CO-oxidation reaction in the atmosphere, which allows for a simplistic albeit initial consideration of photochemistry in the model. 

The methods for using PACMAN and this study differ in the parameters and their respective parameter space explored for each of the TRAPPIST-1 planets. While PACMAN employed a Monte Carlo approach to explore parameter space from many parameters, this study focused on the end points of the parameter space with a nominal case within that range being explored. These two parameters were oxygen fugacity (fO$_2$) and water mass fraction (WMF) for this study. However, a thorough grid search exploring parameter space to validate the model for Earth was done - specifically exploring parameter space for initial mantle temperature and initial fraction of water beginning in the mantle of the planets. This helped set the initial values for the TRAPPIST-1 study. It was crucial to validate the model for Earth before using similar values for the TRAPPIST-1 planets. Lastly, PACMAN was used to study all TRAPPIST-1 planets, while here we focus on planets d, e and f. 

In \citeA{krissansen-totton2022}, they used Monte Carlo simulations across 26 free parameters, including the initial volatile budget, and found that anoxic modern day atmospheres were the most probable outcomes for planets e and f. Their median case for TRAPPIST-1e finds slightly higher H$_2$O partial pressures and substantially higher CO$_2$ pressures than our nominal case at IW+4 at the present day. While it is difficult to determine why our CO$_2$ abundances differ, plausible explanations include differences in the continental weathering rate models used or the higher mantle fO$_2$ that they find as a result of oxidation by escape of hydrogen from an early steam atmosphere. Their model calculates mantle fO$_2$ as a result of reactions with atmospheric oxygen, and they find a nominal fO$_2$ higher than modern Earth, around IW+5.5 (QFM+1.5), but with a range that spans from IW – IW+8. For TRAPPIST-1d, they find a nominal fO$_2$ above the modern Earth, at $\sim$IW+5.5 (QFM+1.5) at the end of the simulation, with a huge range of possible fO$_2$ from IW+1--IW+9. Our nominal case for TRAPPIST-1d at IW+4 is consistent with their median atmospheric pressures, with water vapor dominant and less abundant CO$_2$.  In our results for planet d, we see that water vapor is the most dominant gas in the atmosphere even when varying oxygen fugacity (fO$_2$) and water mass fraction (WMF). This, however, could be due to the partial pressure calculation for water being the saturated vapor pressure. The results from planet f seem to differ the most from the results with PACMAN, possibly due to the suppression of outgassing by pressure noted by \citeA{krissansen-totton2021waterworlds}, which we do not include. They find TRAPPIST-1f to be relatively reduced compared to the inner planets, with nominal fO$_2$ of IW+4.4, with a range from IW to IW+7. They find CO$_2$ to dominate the atmosphere at present day, with minor water vapor, whereas we find CO to be dominant, even for our highest fO$_2$ case, due to very strong carbon cycling that draws CO$_2$ out of the atmosphere.  Water is mostly present on planet f on its surface as an ocean and remains the highest for all three planets in this study regardless of oxygen fugacity.

In PACMAN-P \cite{krissansen-totton2024erosion}, the PACMAN model has been upgraded to include an earlier primary (H$_2$) atmosphere, as well as the addition of other gases, including H$_2$, CO, and CH$_4$. For their nominal TRAPPIST-1e case with no primary atmosphere, they find mantle fO$_2$ comparable to present day Earth, and large present day atmospheres composed mostly of CO$_2$ and O$_2$, which is different from the low atmospheric pressures we find for our nominal case. For the case with a primary atmosphere, they predict a much more reduced mantle fO$_2$, around IW+1, and find a thick ocean with an atmosphere with 10s of bars of CO$_2$. In contrast, our TRAPPIST-1e model with an fO$_2$ of IW predicts low atmospheric pressures of 10$^{-4}$ bars, dominated by water vapor and H$_2$. In general, our model predicts much lower total atmospheric pressures than PACMAN produces, without including hydrodynamic escape. This is potentially due to the assumed efficiencies of silicate weathering, ocean formation, and melt production adopted by the two different models. 

More recently, \citeA{thomas2025} aimed to provide statistical geochemical constraints on the outgassing of water, H$_2$, CH$_4$, CO, and CO$_2$ for each of the TRAPPIST-1 planets. The model is more simple than PACMAN, but provides valuable insights on the volatile outgassing rates of the TRAPPIST-1 planets. Their model samples a range of water content of the bulk mantle and then calculates the partitioning of water into magma during partial melting. Their model then samples parameter space for the rate at which this magma is deposited onto the surface and lastly calculates the volatile outgassing rate using an equilibrium magma outgassing chemistry model. There is agreement between our results and their results for their outgassing rates for the various gas species we track. They found outgassing rates of 10$^{8}$--10$^{12}$ kg/yr for water, 10$^4$--10$^{11}$ kg/yr for H$_2$, 10$^3$ – 10$^{12}$ kg/yr for CO, and 10$^5$-–10$^{13}$ kg/yr for CO$_2$. They also considered the effect of tidal heating on outgassing rates, which we do not consider here. Overall while our study shows the evolution of the outgassing rates over time and how they vary depending on the initial oxygen fugacity, WMF and distance from the star, there is general agreement on the outgassing rates for the gas species explored here.

\section{Conclusion}

In this paper, we made significant improvements to the deep water cycle model set forth in \citeA{schaefer2015persistence} and added the carbon cycle, which allowed us to add three more gas species to the model - CO$_2$, CO and H$_2$ in addition to water (H$_2$O). We added crucial calculations to the model such as surface temperature, mantle oxygen fugacity, atmospheric oxidation reactions and diffusion-limited escape of H$_2$. We validated the model for Earth and then we used it to explore the TRAPPIST-1 planets in and near the habitable zone - planets d, e and f. The two major aspects we wanted to test in this study were the effects of oxygen fugacity and of initial water abundance on the outgassing rates and subsequent partial pressures of the gases mentioned during a planet's geologic evolution. We find that oxygen fugacity exerts a considerably weaker influence on the atmospheric composition and total pressure than suggested by models like \citeA{ortenzi2020mantle}, owing to their omission of geologic sinks for key gases. The final atmosphere compositions are relatively similar between IW and IW-2 for both planet e and f, with more abundant H$_2$ gas, but atmosphere compositions at IW+4 (similar to modern Earth) show more pronounced differences with the other fO$_2$ cases, with more abundant CO$_2$ and H$_2$O. For TRAPPIST-1d, the nominal volatile abundances are so low that there are only minor differences in atmosphere composition across the whole range of fO$_2$ that we explored.

We also studied the effect of initial water abundance as water mass fraction (WMF) on the TRAPPIST-1 planets by keeping the nominal oxygen fugacity case (IW+4). With increased WMF, CO$_2$, H$_2$ and H$_2$O dominate the atmosphere through most of its late evolution into the present day for all planets. However, we saw a bigger effect of the WMF on the inner two planets in this study planets d and e, with more of the gas species present and dominating the atmosphere compared to its lower WMF counterparts that only had two or three gas species at much lower partial pressures. Planet f on the other hand showed an increase in CO abundance at present day for a lower WMF, while exhibiting higher abundances of the other gas species in the higher WMF case. 

Our study highlights the importance of oxygen fugacity and volatile budget on the atmosphere evolution of the TRAPPIST-1 planets, but these models are applicable to other rocky exoplanets as well. In future work, we will extend this model to include other volatile elements, including nitrogen and sulfur, and include additional tectonic processes, including a magma ocean stage. Additionally, incorporating more atmospheric chemical reactions is an important but challenging task due to differences in timescale between atmospheric and tectonic processes, but these additions would help us further understand the abundances of more minor and trace gas species in the atmospheres of rocky exoplanets. 

\section{Open Research}

The code input files, model scripts, the output files, and the jupyter notebook used for modeling presented in this paper are available at Zenodo via https://doi.org/10.5281/zenodo.17544180 \cite{https://doi.org/10.5281/zenodo.17544180}.

\acknowledgments

J.P. thanks Prof. Sabine Stanley for their support during this work. J.P. also thanks the CHAMPs team (Consortium on Habitability and Atmospheres of M-dwarf Planets) for enabling this study. This material is based upon work performed by J.P., L.S., E.S., K.B.S., H.C., and J.L. as part of the CHAMPs team, supported by the National Aeronautics and Space Administration (NASA) under Grant Nos.80NSSC21K0905 and 80NSSC23K1399 issued through the Interdisciplinary Consortia for Astrobiology Research (ICAR) program. This work was also carried out through the support of the NSF Graduate Research Fellowship (JHU award No.DGE-1746891) and by the Maryland Space Grant Consortium Graduate Fellowship. J.P. acknowledges her support system including family, friends, and welcoming spaces such as LUMA and Astrobites, which were crucial in supporting her to finish this work.

\bibliography{references}

@ARTICLE{reyle_gaia,
       author = {{Reyl{\'e}}, C. and {Jardine}, K. and {Fouqu{\'e}}, P. and {Caballero}, J.~A. and {Smart}, R.~L. and {Sozzetti}, A.},
        title = "{The 10 parsec sample in the Gaia era}",
      journal = {Astronomy I\& Astrophysics},
         year = 2021,
        month = jun,
       volume = {650},
          eid = {A201},
        pages = {A201},
          doi = {10.1051/0004-6361/202140985},
archivePrefix = {arXiv},
       eprint = {2104.14972},
 primaryClass = {astro-ph.SR},
       adsurl = {https://ui.adsabs.harvard.edu/abs/2021A&A...650A.201R}
}

@ARTICLE{dressing_occurrence,
       author = {{Dressing}, Courtney D. and {Charbonneau}, David},
        title = "{The Occurrence of Potentially Habitable Planets Orbiting M Dwarfs Estimated from the Full Kepler Dataset and an Empirical Measurement of the Detection Sensitivity}",
      journal = {The Astrophysical Journal},
         year = 2015,
        month = jul,
       volume = {807},
       number = {1},
          eid = {45},
        pages = {45},
          doi = {10.1088/0004-637X/807/1/45},
archivePrefix = {arXiv},
       eprint = {1501.01623},
 primaryClass = {astro-ph.EP},
       adsurl = {https://ui.adsabs.harvard.edu/abs/2015ApJ...807...45D}
}

@article{ricker_tess,
author = {George R. Ricker and Joshua N. Winn and Roland Vanderspek and David W. Latham and G{\'a}sp{\'a}r  {\'A}. Bakos and Jacob L. Bean and Zachory K. Berta-Thompson and Timothy M. Brown and Lars Buchhave and Nathaniel R. Butler and R. Paul Butler and William J. Chaplin and David B. Charbonneau and J{\o}rgen Christensen-Dalsgaard and Mark Clampin and Drake Deming and John P. Doty and Nathan De Lee and Courtney Dressing and Edward W. Dunham and Michael Endl and Fran{\c{c}}ois Fressin and Jian Ge and Thomas Henning and Matthew J. Holman and Andrew W. Howard and Shigeru Ida and Jon M. Jenkins and Garrett Jernigan and John Asher Johnson and Lisa Kaltenegger and Nobuyuki Kawai and Hans Kjeldsen and Gregory Laughlin and Alan M. Levine and Douglas Lin and Jack J. Lissauer and Phillip MacQueen and Geoffrey Marcy and Peter R. McCullough and Timothy D. Morton and Norio Narita and Martin Paegert and Enric Palle and Francesco Pepe and Joshua Pepper and Andreas Quirrenbach and Stephen A. Rinehart and Dimitar Sasselov and Bun’ei Sato and Sara Seager and Alessandro Sozzetti and Keivan G. Stassun and Peter Sullivan and Andrew Szentgyorgyi and Guillermo Torres and Stephane Udry and Joel Villasenor},
title = {{Transiting Exoplanet Survey Satellite}},
volume = {1},
journal = {Journal of Astronomical Telescopes, Instruments, and Systems},
number = {1},
publisher = {SPIE},
pages = {014003},
year = {2014},
doi = {10.1117/1.JATIS.1.1.014003},
URL = {https://doi.org/10.1117/1.JATIS.1.1.014003}
}

@ARTICLE{agol_2021,
       author = {{Agol}, Eric and {Dorn}, Caroline and {Grimm}, Simon L. and {Turbet}, Martin and {Ducrot}, Elsa and {Delrez}, Laetitia and {Gillon}, Micha{\"e}l and {Demory}, Brice-Olivier and {Burdanov}, Artem and {Barkaoui}, Khalid and {Benkhaldoun}, Zouhair and {Bolmont}, Emeline and {Burgasser}, Adam and {Carey}, Sean and {de Wit}, Julien and {Fabrycky}, Daniel and {Foreman-Mackey}, Daniel and {Haldemann}, Jonas and {Hernandez}, David M. and {Ingalls}, James and {Jehin}, Emmanuel and {Langford}, Zachary and {Leconte}, J{\'e}r{\'e}my and {Lederer}, Susan M. and {Luger}, Rodrigo and {Malhotra}, Renu and {Meadows}, Victoria S. and {Morris}, Brett M. and {Pozuelos}, Francisco J. and {Queloz}, Didier and {Raymond}, Sean N. and {Selsis}, Franck and {Sestovic}, Marko and {Triaud}, Amaury H.~M.~J. and {Van Grootel}, Valerie},
        title = "{Refining the Transit-timing and Photometric Analysis of TRAPPIST-1: Masses, Radii, Densities, Dynamics, and Ephemerides}",
      journal = {Planetary Science Journal},
         year = 2021,
        month = feb,
       volume = {2},
       number = {1},
          eid = {1},
        pages = {1},
          doi = {10.3847/PSJ/abd022},
archivePrefix = {arXiv},
       eprint = {2010.01074},
 primaryClass = {astro-ph.EP},
       adsurl = {https://ui.adsabs.harvard.edu/abs/2021PSJ.....2....1A}
}

@article{schaefer2015persistence,
  title={The persistence of oceans on Earth-like planets: Insights from the deep-water cycle},
  author={Schaefer, Laura and Sasselov, Dimitar},
  journal={The Astrophysical Journal},
  volume={801},
  number={1},
  pages={40},
  year={2015},
  publisher={IOP Publishing}
}

@article{holloway1992high,
  title={High-pressure fluid-absent melting experiments in the presence of graphite: oxygen fugacity, ferric/ferrous ratio and dissolved CO2},
  author={Holloway, John R and Pan, Vivian and Gudmundsson, Gisli},
  journal={European Journal of Mineralogy},
  volume={4},
  number={1},
  pages={105--114},
  year={1992}
}

@article{GROTT2011,
title = {Volcanic outgassing of CO2 and H2O on Mars},
journal = {Earth and Planetary Science Letters},
volume = {308},
number = {3},
pages = {391-400},
year = {2011},
issn = {0012-821X},
doi = {https://doi.org/10.1016/j.epsl.2011.06.014},
url = {https://www.sciencedirect.com/science/article/pii/S0012821X11003736},
author = {M. Grott and A. Morschhauser and D. Breuer and E. Hauber}
}

@ARTICLE{foley2015,
       author = {{Foley}, Bradford J.},
        title = "{The Role of Plate Tectonic-Climate Coupling and Exposed Land Area in the Development of Habitable Climates on Rocky Planets}",
      journal = {The Astrophysical Journal},
         year = 2015,
        month = oct,
       volume = {812},
       number = {1},
          eid = {36},
        pages = {36},
          doi = {10.1088/0004-637X/812/1/36},
archivePrefix = {arXiv},
       eprint = {1509.00427},
 primaryClass = {astro-ph.EP},
       adsurl = {https://ui.adsabs.harvard.edu/abs/2015ApJ...812...36F}
}

@article{zahnle2008photochemical,
  title={Photochemical instability of the ancient Martian atmosphere},
  author={Zahnle, Kevin and Haberle, Robert M and Catling, David C and Kasting, James F},
  journal={Journal of Geophysical Research: Planets},
  volume={113},
  number={E11},
  year={2008},
  publisher={Wiley Online Library}
}

@book{fegley2012practical,
  title={Practical chemical thermodynamics for geoscientists},
  author={Fegley Jr, Bruce},
  year={2012},
  publisher={Academic Press}
}

@ARTICLE{driscoll2013,
       author = {{Driscoll}, P. and {Bercovici}, D.},
        title = "{Divergent evolution of Earth and Venus: Influence of degassing, tectonics, and magnetic fields}",
      journal = {Icarus},
         year = 2013,
       volume = {226},
       number = {2},
        pages = {1447-1464},
          doi = {10.1016/j.icarus.2013.07.025},
       adsurl = {https://ui.adsabs.harvard.edu/abs/2013Icar..226.1447D}
}

@ARTICLE{Schaefer2017redox,
       author = {{Schaefer}, Laura and {Fegley}, Bruce, Jr.},
        title = "{Redox States of Initial Atmospheres Outgassed on Rocky Planets and Planetesimals}",
      journal = {The Astrophysical Journal},
         year = 2017,
        month = jul,
       volume = {843},
       number = {2},
          eid = {120},
        pages = {120},
          doi = {10.3847/1538-4357/aa784f},
       adsurl = {https://ui.adsabs.harvard.edu/abs/2017ApJ...843..120S}
}

@ARTICLE{Schaefer2017core,
       author = {{Schaefer}, Laura and {Jacobsen}, Stein B. and {Remo}, John L. and {Petaev}, M.~I. and {Sasselov}, Dimitar D.},
        title = "{Metal-silicate Partitioning and Its Role in Core Formation and Composition on Super-Earths}",
      journal = {The Astrophysical Journal},
         year = 2017,
        month = feb,
       volume = {835},
       number = {2},
          eid = {234},
        pages = {234},
          doi = {10.3847/1538-4357/835/2/234},
       adsurl = {https://ui.adsabs.harvard.edu/abs/2017ApJ...835..234S}
}

@ARTICLE{elkinstanton2008,
       author = {{Elkins-Tanton}, L.~T.},
        title = "{Linked magma ocean solidification and atmospheric growth for Earth and Mars}",
      journal = {Earth and Planetary Science Letters},
         year = 2008,
        month = jul,
       volume = {271},
       number = {1-4},
        pages = {181-191},
          doi = {10.1016/j.epsl.2008.03.062},
       adsurl = {https://ui.adsabs.harvard.edu/abs/2008E&PSL.271..181E}
}

@ARTICLE{rubie2015,
       author = {{Rubie}, D.~C. and {Jacobson}, S.~A. and {Morbidelli}, A. and {O'Brien}, D.~P. and {Young}, E.~D. and {de Vries}, J. and {Nimmo}, F. and {Palme}, H. and {Frost}, D.~J.},
        title = "{Accretion and differentiation of the terrestrial planets with implications for the compositions of early-formed Solar System bodies and accretion of water}",
      journal = {Icarus},
         year = 2015,
        month = mar,
       volume = {248},
        pages = {89-108},
          doi = {10.1016/j.icarus.2014.10.015},
archivePrefix = {arXiv},
       eprint = {1410.3509},
 primaryClass = {astro-ph.EP},
       adsurl = {https://ui.adsabs.harvard.edu/abs/2015Icar..248...89R}
}

@article{zahnle2010,
  title={Earth’s earliest atmospheres},
  author={Zahnle, Kevin and Schaefer, Laura and Fegley, Bruce},
  journal={Cold Spring Harbor perspectives in biology},
  volume={2},
  number={10},
  pages={a004895},
  year={2010},
  publisher={Cold Spring Harbor Lab}
}

@article{baulch_1992,
	title = {Evaluated {Kinetic} {Data} for {Combustion} {Modelling}},
	volume = {21},
	issn = {0047-2689},
	url = {https://doi.org/10.1063/1.555908},
	doi = {10.1063/1.555908},
	number = {3},
	journal = {Journal of Physical and Chemical Reference Data},
	author = {Baulch, D. L. and Cobos, C. J. and Cox, R. A. and Esser, C. and Frank, P. and Just, Th. and Kerr, J. A. and Pilling, M. J. and Troe, J. and Walker, R. W. and Warnatz, J.},
	month = may,
	year = {1992},
	note = {\_eprint: https://pubs.aip.org/aip/jpr/article-pdf/21/3/411/17929929/411\_1\_1.555908.pdf},
	pages = {411--411},
}

@article{sandu_2011,
author = {Sandu, Constantin and Lenardic, Adrian and McGovern, Patrick},
title = {The effects of deep water cycling on planetary thermal evolution},
journal = {Journal of Geophysical Research: Solid Earth},
volume = {116},
number = {B12},
pages = {},
doi = {https://doi.org/10.1029/2011JB008405},
url = {https://agupubs.onlinelibrary.wiley.com/doi/abs/10.1029/2011JB008405},
eprint = {https://agupubs.onlinelibrary.wiley.com/doi/pdf/10.1029/2011JB008405},
year = {2011}
}

@ARTICLE{gillon2017,
       author = {{Gillon}, Micha{\"e}l and {Triaud}, Amaury H.~M.~J. and {Demory}, Brice-Olivier and {Jehin}, Emmanu{\"e}l and {Agol}, Eric and {Deck}, Katherine M. and {Lederer}, Susan M. and {de Wit}, Julien and {Burdanov}, Artem and {Ingalls}, James G. and {Bolmont}, Emeline and {Leconte}, Jeremy and {Raymond}, Sean N. and {Selsis}, Franck and {Turbet}, Martin and {Barkaoui}, Khalid and {Burgasser}, Adam and {Burleigh}, Matthew R. and {Carey}, Sean J. and {Chaushev}, Aleksander and {Copperwheat}, Chris M. and {Delrez}, Laetitia and {Fernandes}, Catarina S. and {Holdsworth}, Daniel L. and {Kotze}, Enrico J. and {Van Grootel}, Val{\'e}rie and {Almleaky}, Yaseen and {Benkhaldoun}, Zouhair and {Magain}, Pierre and {Queloz}, Didier},
        title = "{Seven temperate terrestrial planets around the nearby ultracool dwarf star TRAPPIST-1}",
      journal = {Nature},
         year = 2017,
        month = feb,
       volume = {542},
       number = {7642},
        pages = {456-460},
          doi = {10.1038/nature21360},
archivePrefix = {arXiv},
       eprint = {1703.01424},
 primaryClass = {astro-ph.EP},
       adsurl = {https://ui.adsabs.harvard.edu/abs/2017Natur.542..456G}
}

@ARTICLE{greene2023,
       author = {{Greene}, Thomas P. and {Bell}, Taylor J. and {Ducrot}, Elsa and {Dyrek}, Achr{\`e}ne and {Lagage}, Pierre-Olivier and {Fortney}, Jonathan J.},
        title = "{Thermal emission from the Earth-sized exoplanet TRAPPIST-1 b using JWST}",
      journal = {Nature},
         year = 2023,
        month = jun,
       volume = {618},
       number = {7963},
        pages = {39-42},
          doi = {10.1038/s41586-023-05951-7},
archivePrefix = {arXiv},
       eprint = {2303.14849},
 primaryClass = {astro-ph.EP},
       adsurl = {https://ui.adsabs.harvard.edu/abs/2023Natur.618...39G}
}

@ARTICLE{may2023,
       author = {{May}, E.~M. and {MacDonald}, Ryan J. and {Bennett}, Katherine A. and {Moran}, Sarah E. and {Wakeford}, Hannah R. and {Peacock}, Sarah and {Lustig-Yaeger}, Jacob and {Highland}, Alicia N. and {Stevenson}, Kevin B. and {Sing}, David K. and {Mayorga}, L.~C. and {Batalha}, Natasha E. and {Kirk}, James and {L{\'o}pez-Morales}, Mercedes and {Valenti}, Jeff A. and {Alam}, Munazza K. and {Alderson}, Lili and {Fu}, Guangwei and {Gonzalez-Quiles}, Junellie and {Lothringer}, Joshua D. and {Rustamkulov}, Zafar and {Sotzen}, Kristin S.},
        title = "{Double Trouble: Two Transits of the Super-Earth GJ 1132 b Observed with JWST NIRSpec G395H}",
      journal = {The Astrophysical Journal Letters},
         year = 2023,
        month = dec,
       volume = {959},
       number = {1},
          eid = {L9},
        pages = {L9},
          doi = {10.3847/2041-8213/ad054f},
archivePrefix = {arXiv},
       eprint = {2310.10711},
 primaryClass = {astro-ph.EP},
       adsurl = {https://ui.adsabs.harvard.edu/abs/2023ApJ...959L...9M}
}

@ARTICLE{kirk2024,
       author = {{Kirk}, James and {Stevenson}, Kevin B. and {Fu}, Guangwei and {Lustig-Yaeger}, Jacob and {Moran}, Sarah E. and {Peacock}, Sarah and {Alam}, Munazza K. and {Batalha}, Natasha E. and {Bennett}, Katherine A. and {Gonzalez-Quiles}, Junellie and {L{\'o}pez-Morales}, Mercedes and {Lothringer}, Joshua D. and {MacDonald}, Ryan J. and {May}, E.~M. and {Mayorga}, L.~C. and {Rustamkulov}, Zafar and {Sing}, David K. and {Sotzen}, Kristin S. and {Valenti}, Jeff A. and {Wakeford}, Hannah R.},
        title = "{JWST/NIRCam Transmission Spectroscopy of the Nearby Sub-Earth GJ 341b}",
      journal = {The Astronomical Journal},
         year = 2024,
        month = mar,
       volume = {167},
       number = {3},
          eid = {90},
        pages = {90},
          doi = {10.3847/1538-3881/ad19df},
archivePrefix = {arXiv},
       eprint = {2401.06043},
 primaryClass = {astro-ph.EP},
       adsurl = {https://ui.adsabs.harvard.edu/abs/2024AJ....167...90K}
}

@ARTICLE{moran2023,
       author = {{Moran}, Sarah E. and {Stevenson}, Kevin B. and {Sing}, David K. and {MacDonald}, Ryan J. and {Kirk}, James and {Lustig-Yaeger}, Jacob and {Peacock}, Sarah and {Mayorga}, L.~C. and {Bennett}, Katherine A. and {L{\'o}pez-Morales}, Mercedes and {May}, E.~M. and {Rustamkulov}, Zafar and {Valenti}, Jeff A. and {Adams Redai}, J{\'e}a I. and {Alam}, Munazza K. and {Batalha}, Natasha E. and {Fu}, Guangwei and {Gonzalez-Quiles}, Junellie and {Highland}, Alicia N. and {Kruse}, Ethan and {Lothringer}, Joshua D. and {Ortiz Ceballos}, Kevin N. and {Sotzen}, Kristin S. and {Wakeford}, Hannah R.},
        title = "{High Tide or Riptide on the Cosmic Shoreline? A Water-rich Atmosphere or Stellar Contamination for the Warm Super-Earth GJ 486b from JWST Observations}",
      journal = {The Astrophysical Journal Letters},
         year = 2023,
        month = may,
       volume = {948},
       number = {1},
          eid = {L11},
        pages = {L11},
          doi = {10.3847/2041-8213/accb9c},
archivePrefix = {arXiv},
       eprint = {2305.00868},
 primaryClass = {astro-ph.EP},
       adsurl = {https://ui.adsabs.harvard.edu/abs/2023ApJ...948L..11M}
}

@ARTICLE{lustig-yager2023,
       author = {{Lustig-Yaeger}, Jacob and {Fu}, Guangwei and {May}, E.~M. and {Ceballos}, Kevin N. Ortiz and {Moran}, Sarah E. and {Peacock}, Sarah and {Stevenson}, Kevin B. and {Kirk}, James and {L{\'o}pez-Morales}, Mercedes and {MacDonald}, Ryan J. and {Mayorga}, L.~C. and {Sing}, David K. and {Sotzen}, Kristin S. and {Valenti}, Jeff A. and {Redai}, J{\'e}a I. Adams and {Alam}, Munazza K. and {Batalha}, Natasha E. and {Bennett}, Katherine A. and {Gonzalez-Quiles}, Junellie and {Kruse}, Ethan and {Lothringer}, Joshua D. and {Rustamkulov}, Zafar and {Wakeford}, Hannah R.},
        title = "{A JWST transmission spectrum of the nearby Earth-sized exoplanet LHS 475 b}",
      journal = {Nature Astronomy},
         year = 2023,
        month = nov,
       volume = {7},
        pages = {1317-1328},
          doi = {10.1038/s41550-023-02064-z},
archivePrefix = {arXiv},
       eprint = {2301.04191},
 primaryClass = {astro-ph.EP},
       adsurl = {https://ui.adsabs.harvard.edu/abs/2023NatAs...7.1317L}
}

@ARTICLE{zieba2023,
       author = {{Zieba}, Sebastian and {Kreidberg}, Laura and {Ducrot}, Elsa and {Gillon}, Micha{\"e}l and {Morley}, Caroline and {Schaefer}, Laura and {Tamburo}, Patrick and {Koll}, Daniel D.~B. and {Lyu}, Xintong and {Acu{\~n}a}, Lorena and {Agol}, Eric and {Iyer}, Aishwarya R. and {Hu}, Renyu and {Lincowski}, Andrew P. and {Meadows}, Victoria S. and {Selsis}, Franck and {Bolmont}, Emeline and {Mandell}, Avi M. and {Suissa}, Gabrielle},
        title = "{No thick carbon dioxide atmosphere on the rocky exoplanet TRAPPIST-1 c}",
      journal = {Nature},
         year = 2023,
        month = aug,
       volume = {620},
       number = {7975},
        pages = {746-749},
          doi = {10.1038/s41586-023-06232-z},
archivePrefix = {arXiv},
       eprint = {2306.10150},
 primaryClass = {astro-ph.EP},
       adsurl = {https://ui.adsabs.harvard.edu/abs/2023Natur.620..746Z}
}

@INPROCEEDINGS{lichtenberg2023,
       author = {{Lichtenberg}, Tim and {Schaefer}, Laura K. and {Nakajima}, Miki and {Fischer}, Rebecca A.},
        title = "{Geophysical Evolution During Rocky Planet Formation}",
    booktitle = {Protostars and Planets VII},
         year = 2023,
       editor = {{Inutsuka}, S. and {Aikawa}, Y. and {Muto}, T. and {Tomida}, K. and {Tamura}, M.},
       series = {Astronomical Society of the Pacific Conference Series},
       volume = {534},
        month = jul,
        pages = {907},
          doi = {10.48550/arXiv.2203.10023},
archivePrefix = {arXiv},
       eprint = {2203.10023},
 primaryClass = {astro-ph.EP},
       adsurl = {https://ui.adsabs.harvard.edu/abs/2023ASPC..534..907L}
}

@article{ortenzi2020mantle,
  title={Mantle redox state drives outgassing chemistry and atmospheric composition of rocky planets},
  author={Ortenzi, G and Noack, L and Sohl, F and Guimond, CM and Grenfell, JL and Dorn, C and Schmidt, JM and Vulpius, S and Katyal, N and Kitzmann, D and others},
  journal={Scientific reports},
  volume={10},
  number={1},
  pages={10907},
  year={2020},
  publisher={Nature Publishing Group UK London}
}

@ARTICLE{gaillard2014,
       author = {{Gaillard}, Fabrice and {Scaillet}, Bruno},
        title = "{A theoretical framework for volcanic degassing chemistry in a comparative planetology perspective and implications for planetary atmospheres}",
      journal = {Earth and Planetary Science Letters},
         year = 2014,
        month = oct,
       volume = {403},
        pages = {307-316},
          doi = {10.1016/j.epsl.2014.07.009},
       adsurl = {https://ui.adsabs.harvard.edu/abs/2014E&PSL.403..307G}
}

@ARTICLE{hirschmann2012,
       author = {{Hirschmann}, Marc M.},
        title = "{Magma ocean influence on early atmosphere mass and composition}",
      journal = {Earth and Planetary Science Letters},
         year = 2012,
        month = aug,
       volume = {341},
        pages = {48-57},
          doi = {10.1016/j.epsl.2012.06.015},
       adsurl = {https://ui.adsabs.harvard.edu/abs/2012E&PSL.341...48H}
}

@article{wadhwa2008,
  title={Redox conditions on small bodies, the Moon and Mars},
  author={Wadhwa, Meenakshi},
  journal={Reviews in Mineralogy and Geochemistry},
  volume={68},
  number={1},
  pages={493--510},
  year={2008},
  publisher={Mineralogical Society of America}
}

@article{nicklas2021,
  title={Uniform oxygen fugacity of shergottite mantle sources and an oxidized martian lithosphere},
  author={Nicklas, Robert W and Day, James MD and Vaci, Zoltan and Udry, Arya and Liu, Yang and Tait, Kimberly T},
  journal={Earth and Planetary Science Letters},
  volume={564},
  pages={116876},
  year={2021},
  publisher={Elsevier}
}

@book{catling2017atmospheric,
  title={Atmospheric evolution on inhabited and lifeless worlds},
  author={Catling, David C and Kasting, James F},
  year={2017},
  publisher={Cambridge University Press}
}

@incollection{jaupart2015treatise,
title = {7.06 - Temperatures, Heat, and Energy in the Mantle of the Earth},
editor = {Gerald Schubert},
booktitle = {Treatise on Geophysics (Second Edition)},
publisher = {Elsevier},
edition = {Second Edition},
address = {Oxford},
pages = {223-270},
year = {2015},
isbn = {978-0-444-53803-1},
doi = {https://doi.org/10.1016/B978-0-444-53802-4.00126-3},
url = {https://www.sciencedirect.com/science/article/pii/B9780444538024001263},
author = {C. Jaupart and S. Labrosse and F. Lucazeau and J.-C. Mareschal}
}

@article{peslier2017water,
  title={Water in the Earth’s interior: distribution and origin},
  author={Peslier, Anne H and Sch{\"o}nb{\"a}chler, Maria and Busemann, Henner and Karato, Shun-Ichiro},
  journal={Space Science Reviews},
  volume={212},
  pages={743--810},
  year={2017},
  publisher={Springer}
}

@article{hirschmann2018comparative,
  title={Comparative deep Earth volatile cycles: The case for C recycling from exosphere/mantle fractionation of major (H2O, C, N) volatiles and from H2O/Ce, CO2/Ba, and CO2/Nb exosphere ratios},
  author={Hirschmann, Marc M},
  journal={Earth and Planetary Science Letters},
  volume={502},
  pages={262--273},
  year={2018},
  publisher={Elsevier}
}

@article{mccubbin2019origin,
  title={Origin and abundances of H2O in the terrestrial planets, Moon, and asteroids},
  author={McCubbin, Francis M and Barnes, Jessica J},
  journal={Earth and Planetary Science Letters},
  volume={526},
  pages={115771},
  year={2019},
  publisher={Elsevier}
}

@article{ohtani2021hydration,
  title={Hydration and dehydration in Earth's interior},
  author={Ohtani, Eiji},
  journal={Annual Review of Earth and Planetary Sciences},
  volume={49},
  number={1},
  pages={253--278},
  year={2021},
  publisher={Annual Reviews}
}

@article{lodders2009abundances,
  title={4.4 Abundances of the elements in the Solar System},
  author={Lodders, K and Palme, H and Gail, HP},
  journal={Solar system},
  pages={712--770},
  year={2009},
  publisher={Springer}
}

@article{canil1997vanadium,
  title={Vanadium partitioning and the oxidation state of Archaean komatiite magmas},
  author={Canil, Dante},
  journal={Nature},
  volume={389},
  number={6653},
  pages={842--845},
  year={1997},
  publisher={Nature Publishing Group UK London}
}

@article{delano2001redox,
  title={Redox history of the Earth's interior since~ 3900 Ma: implications for prebiotic molecules},
  author={Delano, John W},
  journal={Origins of Life and Evolution of the Biosphere},
  volume={31},
  pages={311--341},
  year={2001},
  publisher={Springer}
}

@article{li2004constancy,
  title={The constancy of upper mantle fO2 through time inferred from V/Sc ratios in basalts},
  author={Li, Zheng-Xue Anser and Lee, Cin-Ty Aeolus},
  journal={Earth and Planetary Science Letters},
  volume={228},
  number={3-4},
  pages={483--493},
  year={2004},
  publisher={Elsevier}
}

@ARTICLE{krissansen-totton2022,
       author = {{Krissansen-Totton}, J. and {Fortney}, J.~J.},
        title = "{Predictions for Observable Atmospheres of Trappist-1 Planets from a Fully Coupled Atmosphere-Interior Evolution Model}",
      journal = {The Astrophysical Journal},
         year = 2022,
        month = jul,
       volume = {933},
       number = {1},
          eid = {115},
        pages = {115},
          doi = {10.3847/1538-4357/ac69cb},
archivePrefix = {arXiv},
       eprint = {2207.04164},
 primaryClass = {astro-ph.EP},
       adsurl = {https://ui.adsabs.harvard.edu/abs/2022ApJ...933..115K}
}

@ARTICLE{boldog2024,
       author = {{Boldog}, {\'A}d{\'a}m and {Dobos}, Vera and {Kiss}, L{\'a}szl{\'o} L. and {van der Perk}, Marijn and {Barr}, Amy C.},
        title = "{Water content of rocky exoplanets in the habitable zone}",
      journal = {Astronomy I\& Astrophysics},
         year = 2024,
        month = jan,
       volume = {681},
          eid = {A109},
        pages = {A109},
          doi = {10.1051/0004-6361/202346988},
archivePrefix = {arXiv},
       eprint = {2312.01893},
 primaryClass = {astro-ph.EP},
       adsurl = {https://ui.adsabs.harvard.edu/abs/2024A&A...681A.109B}
}

@ARTICLE{thomas2025,
       author = {{Thomas}, Trent B. and {Meadows}, Victoria S. and {Krissansen-Totton}, Joshua and {Gialluca}, Megan T. and {Wogan}, Nicholas F. and {Catling}, David C.},
        title = "{Statistical Geochemical Constraints on Present-day Water Outgassing as a Source of Secondary Atmospheres on the TRAPPIST-1 Exoplanets}",
      journal = {Planetary Science Journal},
         year = 2025,
        month = may,
       volume = {6},
       number = {5},
          eid = {126},
        pages = {126},
          doi = {10.3847/PSJ/add261},
archivePrefix = {arXiv},
       eprint = {2505.03672},
 primaryClass = {astro-ph.EP},
       adsurl = {https://ui.adsabs.harvard.edu/abs/2025PSJ.....6..126T}
}

@article{stern2005evidence,
  title={Evidence from ophiolites, blueschists, and ultrahigh-pressure metamorphic terranes that the modern episode of subduction tectonics began in Neoproterozoic time},
  author={Stern, Robert J},
  journal={Geology},
  volume={33},
  number={7},
  pages={557--560},
  year={2005},
  publisher={Geological Society of America}
}

@article{palin2020secular,
  title={Secular change and the onset of plate tectonics on Earth},
  author={Palin, Richard M and Santosh, M and Cao, Wentao and Li, Shan-Shan and Hern{\'a}ndez-Uribe, David and Parsons, Andrew},
  journal={Earth-Science Reviews},
  volume={207},
  pages={103172},
  year={2020},
  publisher={Elsevier}
}

@article{cawood2018geological,
  title={Geological archive of the onset of plate tectonics},
  author={Cawood, Peter A and Hawkesworth, Chris J and Pisarevsky, Sergei A and Dhuime, Bruno and Capitanio, Fabio A and Nebel, Oliver},
  journal={Philosophical Transactions of the Royal Society A: mathematical, physical and engineering sciences},
  volume={376},
  number={2132},
  pages={20170405},
  year={2018},
  publisher={The Royal Society Publishing}
}

@article{foley2018carbon,
  title={Carbon cycling and habitability of Earth-sized stagnant lid planets},
  author={Foley, Bradford J and Smye, Andrew J},
  journal={Astrobiology},
  volume={18},
  number={7},
  pages={873--896},
  year={2018},
  publisher={Mary Ann Liebert, Inc. 140 Huguenot Street, 3rd Floor New Rochelle, NY 10801 USA}
}

@article{dorn2018outgassing,
  title={Outgassing on stagnant-lid super-Earths},
  author={Dorn, Caroline and Noack, Lena and Rozel, AB},
  journal={Astronomy \& Astrophysics},
  volume={614},
  pages={A18},
  year={2018},
  publisher={EDP Sciences}
}

@article{noack2017volcanism,
  title={Volcanism and outgassing of stagnant-lid planets: implications for the habitable zone},
  author={Noack, Lena and Rivoldini, Attilio and Van Hoolst, Tim},
  journal={Physics of the Earth and Planetary Interiors},
  volume={269},
  pages={40--57},
  year={2017},
  publisher={Elsevier}
}

@article{tosi2017habitability,
  title={The habitability of a stagnant-lid Earth},
  author={Tosi, Nicola and Godolt, Mareike and Stracke, Barbara and Ruedas, Thomas and Grenfell, John Lee and H{\"o}ning, Dennis and Nikolaou, Athanasia and Plesa, A-C and Breuer, Doris and Spohn, Tilman},
  journal={Astronomy \& Astrophysics},
  volume={605},
  pages={A71},
  year={2017},
  publisher={EDP Sciences}
}

@article{sleep2001carbon,
  title={Carbon dioxide cycling and implications for climate on ancient Earth},
  author={Sleep, Norman H and Zahnle, Kevin},
  journal={Journal of Geophysical Research: Planets},
  volume={106},
  number={E1},
  pages={1373--1399},
  year={2001},
  publisher={Wiley Online Library}
}

@article{schaefer2016predictions,
  title={Predictions of the atmospheric composition of GJ 1132b},
  author={Schaefer, Laura and Wordsworth, Robin D and Berta-Thompson, Zachory and Sasselov, Dimitar},
  journal={The Astrophysical Journal},
  volume={829},
  number={2},
  pages={63},
  year={2016},
  publisher={IOP Publishing}
}

@ARTICLE{krissansen-totton2021waterworlds,
       author = {{Krissansen-Totton}, Joshua and {Galloway}, Max L. and {Wogan}, Nicholas and {Dhaliwal}, Jasmeet K. and {Fortney}, Jonathan J.},
        title = "{Waterworlds Probably Do Not Experience Magmatic Outgassing}",
      journal = {The Astrophysical Journal},
         year = 2021,
        month = jun,
       volume = {913},
       number = {2},
          eid = {107},
        pages = {107},
          doi = {10.3847/1538-4357/abf560},
archivePrefix = {arXiv},
       eprint = {2106.08538},
 primaryClass = {astro-ph.EP},
       adsurl = {https://ui.adsabs.harvard.edu/abs/2021ApJ...913..107K}
}

@ARTICLE{krissansen-totton2021venus,
       author = {{Krissansen-Totton}, Joshua and {Fortney}, Jonathan J. and {Nimmo}, Francis},
        title = "{Was Venus Ever Habitable? Constraints from a Coupled Interior-Atmosphere-Redox Evolution Model}",
      journal = {Planetary Science Journal},
         year = 2021,
        month = oct,
       volume = {2},
       number = {5},
          eid = {216},
        pages = {216},
          doi = {10.3847/PSJ/ac2580},
archivePrefix = {arXiv},
       eprint = {2111.00033},
 primaryClass = {astro-ph.EP},
       adsurl = {https://ui.adsabs.harvard.edu/abs/2021PSJ.....2..216K}
}

@ARTICLE{krissansen-totton2023nondetection,
       author = {{Krissansen-Totton}, Joshua},
        title = "{Implications of Atmospheric Nondetections for Trappist-1 Inner Planets on Atmospheric Retention Prospects for Outer Planets}",
      journal = {The Astrophysical Journal Letters},
         year = 2023,
        month = jul,
       volume = {951},
       number = {2},
          eid = {L39},
        pages = {L39},
          doi = {10.3847/2041-8213/acdc26},
archivePrefix = {arXiv},
       eprint = {2306.05397},
 primaryClass = {astro-ph.EP},
       adsurl = {https://ui.adsabs.harvard.edu/abs/2023ApJ...951L..39K}
}

@ARTICLE{krissansen-totton2024erosion,
       author = {{Krissansen-Totton}, Joshua and {Wogan}, Nicholas and {Thompson}, Maggie and {Fortney}, Jonathan J.},
        title = "{The erosion of large primary atmospheres typically leaves behind substantial secondary atmospheres on temperate rocky planets}",
      journal = {Nature Communications},
         year = 2024,
        month = dec,
       volume = {15},
       number = {1},
          eid = {8374},
        pages = {8374},
          doi = {10.1038/s41467-024-52642-6},
archivePrefix = {arXiv},
       eprint = {2409.18940},
 primaryClass = {astro-ph.EP},
       adsurl = {https://ui.adsabs.harvard.edu/abs/2024NatCo..15.8374K}
}

@article{van2024airy,
  title={Airy worlds or barren rocks? On the survivability of secondary atmospheres around the TRAPPIST-1 planets},
  author={Van Looveren, Gwena{\"e}l and G{\"u}del, Manuel and Saikia, Sudeshna Boro and Kislyakova, Kristina},
  journal={Astronomy \& Astrophysics},
  volume={683},
  pages={A153},
  year={2024},
  publisher={EDP Sciences}
}

@article{gialluca2024implications,
  title={The Implications of Thermal Hydrodynamic Atmospheric Escape on the TRAPPIST-1 Planets},
  author={Gialluca, Megan T and Barnes, Rory and Meadows, Victoria S and Garcia, Rodolfo and Birky, Jessica and Agol, Eric},
  journal={The Planetary Science Journal},
  volume={5},
  number={6},
  pages={137},
  year={2024},
  publisher={IOP Publishing}
}

@article{barth2021magma,
  title={Magma ocean evolution of the TRAPPIST-1 planets},
  author={Barth, Patrick and Carone, Ludmila and Barnes, Rory and Noack, Lena and Molli{\`e}re, Paul and Henning, Thomas},
  journal={Astrobiology},
  volume={21},
  number={11},
  pages={1325--1349},
  year={2021},
  publisher={Mary Ann Liebert, Inc., publishers 140 Huguenot Street, 3rd Floor New~…}
}

@article{rice2025uncertainties,
  title={Uncertainties in the Inference of Internal Structure: The Case of TRAPPIST-1 f},
  author={Rice, David R and Huang, Chenliang and Steffen, Jason H and Vazan, Allona},
  journal={The Astrophysical Journal},
  volume={986},
  number={1},
  pages={2},
  year={2025},
  publisher={IOP Publishing}
}

@article{dewit2018atmospheric,
  title={Atmospheric reconnaissance of the habitable-zone Earth-sized planets orbiting TRAPPIST-1},
  author={De Wit, Julien and Wakeford, Hannah R and Lewis, Nikole K and Delrez, Laetitia and Gillon, Micha{\"e}l and Selsis, Frank and Leconte, J{\'e}r{\'e}my and Demory, Brice-Olivier and Bolmont, Emeline and Bourrier, Vincent and others},
  journal={Nature Astronomy},
  volume={2},
  number={3},
  pages={214--219},
  year={2018},
  publisher={Nature Publishing Group UK London}
}

@article{espinoza2025jwst,
  title={JWST-TST DREAMS: NIRSpec/PRISM Transmission Spectroscopy of the Habitable Zone Planet TRAPPIST-1 e},
  author={Espinoza, N{\'e}stor and Allen, Natalie H and Glidden, Ana and Lewis, Nikole K and Seager, Sara and Ca{\~n}as, Caleb I and Grant, David and Gressier, Am{\'e}lie and Courreges, Shelby and Stevenson, Kevin B and others},
  journal={The Astrophysical Journal Letters},
  volume={990},
  number={2},
  pages={L52},
  year={2025},
  publisher={IOP Publishing}
}

@article{glidden2025jwst,
  title={JWST-TST DREAMS: Secondary Atmosphere Constraints for the Habitable Zone Planet TRAPPIST-1 e},
  author={Glidden, Ana and Ranjan, Sukrit and Seager, Sara and Espinoza, N{\'e}stor and MacDonald, Ryan J and Allen, Natalie H and Ca{\~n}as, Caleb I and Grant, David and Gressier, Am{\'e}lie and Stevenson, Kevin B and others},
  journal={The Astrophysical Journal Letters},
  volume={990},
  number={2},
  pages={L53},
  year={2025},
  publisher={IOP Publishing}
}

@article{piaulet2025strict,
  title={Strict limits on potential secondary atmospheres on the temperate rocky exo-Earth TRAPPIST-1 d},
  author={Piaulet-Ghorayeb, Caroline and Benneke, Bj{\"o}rn and Turbet, Martin and Moore, Keavin and Roy, Pierre-Alexis and Lim, Olivia and Doyon, Ren{\'e} and Fauchez, Thomas J and Albert, Lo{\"\i}c and Radica, Michael and others},
  journal={The Astrophysical Journal},
  volume={989},
  number={2},
  pages={181},
  year={2025},
  publisher={IOP Publishing}
}

@article{karato2015water,
  title={Water in the evolution of the Earth and other terrestrial planets},
  author={Karato, SI},
  journal={Treatise on Geophysics},
  volume={9},
  pages={105--144},
  year={2015},
  publisher={Elsevier Amsterdam}
}

@ARTICLE{nutzman2008,
       author = {{Nutzman}, Philip and {Charbonneau}, David},
        title = "{Design Considerations for a Ground-Based Transit Search for Habitable Planets Orbiting M Dwarfs}",
      journal = {Publications of the Astronomical Society of the Pacific},
         year = 2008,
        month = mar,
       volume = {120},
       number = {865},
        pages = {317},
          doi = {10.1086/533420},
archivePrefix = {arXiv},
       eprint = {0709.2879},
 primaryClass = {astro-ph},
       adsurl = {https://ui.adsabs.harvard.edu/abs/2008PASP..120..317N}
}

@ARTICLE{charbonneau2007,
       author = {{Charbonneau}, David and {Deming}, Drake},
        title = "{The Dynamics-Based Approach to Studying Terrestrial Exoplanets}",
      journal = {arXiv e-prints},
         year = 2007,
        month = jun,
          eid = {arXiv:0706.1047},
        pages = {arXiv:0706.1047},
          doi = {10.48550/arXiv.0706.1047},
archivePrefix = {arXiv},
       eprint = {0706.1047},
 primaryClass = {astro-ph},
       adsurl = {https://ui.adsabs.harvard.edu/abs/2007arXiv0706.1047C}
}

@ARTICLE{kopparapu2014,
       author = {{Kopparapu}, Ravi Kumar and {Ramirez}, Ramses M. and {SchottelKotte}, James and {Kasting}, James F. and {Domagal-Goldman}, Shawn and {Eymet}, Vincent},
        title = "{Habitable Zones around Main-sequence Stars: Dependence on Planetary Mass}",
      journal = {The Astrophysical Journal Letters},
         year = 2014,
        month = jun,
       volume = {787},
       number = {2},
          eid = {L29},
        pages = {L29},
          doi = {10.1088/2041-8205/787/2/L29},
archivePrefix = {arXiv},
       eprint = {1404.5292},
 primaryClass = {astro-ph.EP},
       adsurl = {https://ui.adsabs.harvard.edu/abs/2014ApJ...787L..29K}
}

@ARTICLE{bennett2025,
       author = {{Bennett}, Katherine A. and {MacDonald}, Ryan J. and {Peacock}, Sarah and {Perez}, Junellie and {May}, E.~M. and {Moran}, Sarah E. and {Alderson}, Lili and {Lustig-Yaeger}, Jacob and {Wakeford}, Hannah R. and {Sing}, David K. and {Stevenson}, Kevin B. and {Batalha}, Natasha E. and {L{\'o}pez-Morales}, Mercedes and {Alam}, Munazza K. and {Lothringer}, Joshua D. and {Fu}, Guangwei and {Kirk}, James and {Valenti}, Jeff A. and {Mayorga}, L.~C. and {Sotzen}, Kristin S.},
        title = "{Additional JWST/NIRSpec Transits of the Rocky M Dwarf Exoplanet GJ 1132 b Reveal a Featureless Spectrum}",
      journal = {The Astronomical Journal},
         year = 2025,
        month = oct,
       volume = {170},
       number = {4},
          eid = {205},
        pages = {205},
          doi = {10.3847/1538-3881/adf198},
archivePrefix = {arXiv},
       eprint = {2508.10579},
 primaryClass = {astro-ph.EP},
       adsurl = {https://ui.adsabs.harvard.edu/abs/2025AJ....170..205B}
}

@ARTICLE{hirschmann2006,
       author = {{Hirschmann}, Marc M.},
        title = "{Water, Melting, and the Deep Earth H2O Cycle}",
      journal = {Annual Review of Earth and Planetary Sciences},
         year = 2006,
        month = may,
       volume = {34},
        pages = {629-653},
          doi = {10.1146/annurev.earth.34.031405.125211},
       adsurl = {https://ui.adsabs.harvard.edu/abs/2006AREPS..34..629H}
}

@ARTICLE{walker1981,
       author = {{Walker}, J.~C.~G. and {Hays}, P.~B. and {Kasting}, J.~F.},
        title = "{A negative feedback mechanism for the long-term stabilization of the earth's surface temperature}",
      journal = {Journal of Geophysical Research},
         year = 1981,
        month = oct,
       volume = {86},
        pages = {9776-9782},
          doi = {10.1029/JC086iC10p09776},
       adsurl = {https://ui.adsabs.harvard.edu/abs/1981JGR....86.9776W}
}

@ARTICLE{doyle2019,
       author = {{Doyle}, Alexandra E. and {Young}, Edward D. and {Klein}, Beth and {Zuckerman}, Ben and {Schlichting}, Hilke E.},
        title = "{Oxygen fugacities of extrasolar rocks: Evidence for an Earth-like geochemistry of exoplanets}",
      journal = {Science},
         year = 2019,
        month = oct,
       volume = {366},
       number = {6463},
        pages = {356-359},
          doi = {10.1126/science.aax3901},
archivePrefix = {arXiv},
       eprint = {1910.12989},
 primaryClass = {astro-ph.EP},
       adsurl = {https://ui.adsabs.harvard.edu/abs/2019Sci...366..356D}
}

@ARTICLE{guimond2023,
       author = {{Guimond}, Claire Marie and {Shorttle}, Oliver and {Jordan}, Sean and {Rudge}, John F.},
        title = "{A mineralogical reason why all exoplanets cannot be equally oxidizing}",
      journal = {Monthly Notices of the Royal Astronomical Society},
         year = 2023,
        month = nov,
       volume = {525},
       number = {3},
        pages = {3703-3717},
          doi = {10.1093/mnras/stad2486},
archivePrefix = {arXiv},
       eprint = {2308.09505},
 primaryClass = {astro-ph.EP},
       adsurl = {https://ui.adsabs.harvard.edu/abs/2023MNRAS.525.3703G}
}

@ARTICLE{schaefer2017,
       author = {{Schaefer}, Laura and {Fegley}, Jr., Bruce},
        title = "{Redox States of Initial Atmospheres Outgassed on Rocky Planets and Planetesimals}",
      journal = {The Astrophysical Journal},
         year = 2017,
        month = jul,
       volume = {843},
       number = {2},
          eid = {120},
        pages = {120},
          doi = {10.3847/1538-4357/aa784f},
       adsurl = {https://ui.adsabs.harvard.edu/abs/2017ApJ...843..120S}
}

@ARTICLE{kasting1993,
       author = {{Kasting}, James F. and {Eggler}, David H. and {Raeburn}, Stuart P.},
        title = "{Mantle Redox Evolution and the Oxidation State of the Archean Atmosphere}",
      journal = {Journal of Geology},
         year = 1993,
        month = mar,
       volume = {101},
       number = {2},
        pages = {245-257},
          doi = {10.1086/648219},
       adsurl = {https://ui.adsabs.harvard.edu/abs/1993JG....101..245K}
}

@ARTICLE{hirth1996,
       author = {{Hirth}, Greg and {Kohlstedt}, David L.},
        title = "{Water in the oceanic upper mantle: implications for rheology, melt extraction and the evolution of the lithosphere}",
      journal = {Earth and Planetary Science Letters},
         year = 1996,
        month = oct,
       volume = {144},
       number = {1-2},
        pages = {93-108},
          doi = {10.1016/0012-821X(96)00154-9},
       adsurl = {https://ui.adsabs.harvard.edu/abs/1996E&PSL.144...93H}
}

@ARTICLE{karato1990,
       author = {{Karato}, S.},
        title = "{The role of hydrogen in the electrical conductivity of the upper mantle}",
      journal = {Nature},
         year = 1990,
        month = sep,
       volume = {347},
       number = {6290},
        pages = {272-273},
          doi = {10.1038/347272a0},
       adsurl = {https://ui.adsabs.harvard.edu/abs/1990Natur.347..272K}
}

@ARTICLE{karato1993,
       author = {{Karato}, Shun-Ichiro and {Wu}, Patrick},
        title = "{Rheology of the Upper Mantle: A Synthesis}",
      journal = {Science},
         year = 1993,
        month = may,
       volume = {260},
       number = {5109},
        pages = {771-778},
          doi = {10.1126/science.260.5109.771},
       adsurl = {https://ui.adsabs.harvard.edu/abs/1993Sci...260..771K}
}

@ARTICLE{mcgovernshubert1989,
       author = {{McGovern}, Patrick J. and {Schubert}, Gerald},
        title = "{Thermal evolution of the Earth: effects of volatile exchange between atmosphere and interior}",
      journal = {Earth and Planetary Science Letters},
         year = 1989,
        month = dec,
       volume = {96},
       number = {1-2},
        pages = {27-37},
          doi = {10.1016/0012-821X(89)90121-0},
       adsurl = {https://ui.adsabs.harvard.edu/abs/1989E&PSL..96...27M}
}

@ARTICLE{bounama2001,
       author = {{Bounama}, C. and {Franck}, S. and {von Bloh}, W.},
        title = "{The fate of Earth's ocean}",
      journal = {Hydrology and Earth System Sciences},
         year = 2001,
        month = jan,
       volume = {5},
       number = {4},
        pages = {569-576},
          doi = {10.5194/hess-5-569-2001},
       adsurl = {https://ui.adsabs.harvard.edu/abs/2001HESS....5..569B}
}

@ARTICLE{crowley2011,
       author = {{Crowley}, John W. and {G{\'e}rault}, M{\'e}lanie and {O'Connell}, Richard J.},
        title = "{On the relative influence of heat and water transport on planetary dynamics}",
      journal = {Earth and Planetary Science Letters},
         year = 2011,
        month = oct,
       volume = {310},
       number = {3},
        pages = {380-388},
          doi = {10.1016/j.epsl.2011.08.035},
       adsurl = {https://ui.adsabs.harvard.edu/abs/2011E&PSL.310..380C}
}

@ARTICLE{joshkt2017,
       author = {{Krissansen-Totton}, Joshua and {Catling}, David C.},
        title = "{Constraining climate sensitivity and continental versus seafloor weathering using an inverse geological carbon cycle model}",
      journal = {Nature Communications},
         year = 2017,
        month = may,
       volume = {8},
          eid = {15423},
        pages = {15423},
          doi = {10.1038/ncomms15423},
       adsurl = {https://ui.adsabs.harvard.edu/abs/2017NatCo...815423K}
}

@ARTICLE{wolf2025,
       author = {{Wolf}, Eric T. and {Schwieterman}, Edward W. and {Haqq-Misra}, Jacob and {Fauchez}, Thomas J. and {Bastelberger}, Sandra T. and {Leung}, Michaela and {Peacock}, Sarah and {Villanueva}, Geronimo L. and {Kopparapu}, Ravi K.},
        title = "{Chemistry, Climate, and Transmission Spectra of TRAPPIST-1 e Explored with a Multimodel Sparse Sampled Ensemble}",
      journal = {Planetary Science Journal},
         year = 2025,
        month = oct,
       volume = {6},
       number = {10},
          eid = {231},
        pages = {231},
          doi = {10.3847/PSJ/ae031e},
archivePrefix = {arXiv},
       eprint = {2510.18704},
 primaryClass = {astro-ph.EP},
       adsurl = {https://ui.adsabs.harvard.edu/abs/2025PSJ.....6..231W}
}

@ARTICLE{rushby2020,
       author = {{Rushby}, Andrew J. and {Shields}, Aomawa L. and {Wolf}, Eric T. and {Lagu{\"e}}, Marysa and {Burgasser}, Adam},
        title = "{The Effect of Land Albedo on the Climate of Land-dominated Planets in the TRAPPIST-1 System}",
      journal = {The Astrophysical Journal},
         year = 2020,
        month = dec,
       volume = {904},
       number = {2},
          eid = {124},
        pages = {124},
          doi = {10.3847/1538-4357/abbe04},
archivePrefix = {arXiv},
       eprint = {2011.03621},
 primaryClass = {astro-ph.EP},
       adsurl = {https://ui.adsabs.harvard.edu/abs/2020ApJ...904..124R}
}

@ARTICLE{wolf2017,
       author = {{Wolf}, Eric T.},
        title = "{Assessing the Habitability of the TRAPPIST-1 System Using a 3D Climate Model}",
      journal = {The Astrophysical Journal Letters},
         year = 2017,
        month = apr,
       volume = {839},
       number = {1},
          eid = {L1},
        pages = {L1},
          doi = {10.3847/2041-8213/aa693a},
archivePrefix = {arXiv},
       eprint = {1703.05815},
 primaryClass = {astro-ph.EP},
       adsurl = {https://ui.adsabs.harvard.edu/abs/2017ApJ...839L...1W}
}

@ARTICLE{turbet2018,
       author = {{Turbet}, Martin and {Bolmont}, Emeline and {Leconte}, Jeremy and {Forget}, Fran{\c{c}}ois and {Selsis}, Franck and {Tobie}, Gabriel and {Caldas}, Anthony and {Naar}, Joseph and {Gillon}, Micha{\"e}l},
        title = "{Modeling climate diversity, tidal dynamics and the fate of volatiles on TRAPPIST-1 planets}",
      journal = {Astronomy I\& Astrophysics},
         year = 2018,
        month = may,
       volume = {612},
          eid = {A86},
        pages = {A86},
          doi = {10.1051/0004-6361/201731620},
archivePrefix = {arXiv},
       eprint = {1707.06927},
 primaryClass = {astro-ph.EP},
       adsurl = {https://ui.adsabs.harvard.edu/abs/2018A&A...612A..86T}
}

@misc{https://doi.org/10.5281/zenodo.17544180,
doi = {10.5281/zenodo.17544180},
author = {Perez, Junellie and Schaefer, Laura},
title = {Atmosphere-Interior Exchange Model for the TRAPPIST-1 Planets},
publisher = {Zenodo},
year = {2025},
type = {software}
}

\end{document}